\documentclass[12pt]{iopart}

\usepackage{iopams}

\newcommand{\be}{\begin{equation}}
\newcommand{\ee}{\end{equation}}

\newcommand{\ba}{\begin{array}}
\newcommand{\ea}{\end{array}}
\newcommand{\bea}{\begin{eqnarray}}
\newcommand{\eea}{\end{eqnarray}}
\newcommand{\qed}{\begin{flushright} $\square$
                  \end{flushright}
}
\newcommand{\g}{\mathfrak{g}}
\newcommand{\nn}{\nonumber \\}
\newcommand{\prw}{pr \vec{w}_\Q }
\newcommand{\prwc}{pr \vec{w}_C }

\newcommand{\Q}{\mathcal{Q}}
\newtheorem{theorem}{Theorem}
\newtheorem{corollary}{Corollary}
\newtheorem{proposition}{Proposition}
\newcommand{\CP}{\mathbb{C}P^{N-1}}

\begin{document}

\title{Soliton surfaces associated with generalized symmetries of integrable equations}
\author{A M Grundland$^1$ $^2$ and S Post$^1$}
\address{$^1$ Centre de Recherches Math\'ematiques. Universit\'e de Montr\'eal, Montr\'eal CP 6128 (QC) H3C 3J7, Canada}
\address{$^2$ Department of Mathematics and Computer Sciences, Universit\'e du Quebec, Trois-Riviers. CP500 (QC)G9A 5H7, Canada}
\ead{grundlan@crm.umontreal.ca, post@crm.umontreal.ca}

\begin{abstract}In this paper, based on the Fokas, Gel'fand et al approach \cite{FG, FGFL}, we provide a symmetry characterization of continuous deformations of soliton surfaces immersed in a  Lie algebra using the formalism of generalized vector fields, their prolongation structure and  links with the Fr\'echet derivatives. We express the  necessary and sufficient condition for the existence of such surfaces in terms of the invariance criterion for generalized symmetries and identify additional sufficient conditions which admit an explicit integration of the immersion functions of 2D surfaces in Lie algebras. We discuss in detail the $su(N)$-valued immersion functions generated by conformal symmetries of  the $CP^{N-1}$ sigma model defined on either the Minkowski or Euclidean space. We further show that the sufficient conditions for explicit integration of such immersion functions impose additional restrictions on the admissible conformal symmetries of the model defined on Minkowski space. On the other hand,   the sufficient conditions are identically satisfied for arbitrary conformal symmetries of finite action solutions of the $CP^{N-1}$ sigma model defined on Euclidean space.  \end{abstract}


\maketitle
\section{Introduction}
Over the last century, the problem of investigation and  construction of surfaces and their continuous deformations under various types of dynamics has been the subject of extensive research. In particular, the study of surfaces associated with integrable models has been vigorously pursued and  significant results have been obtained concerning the intrinsic geometric properties as well as integrable dynamics of soliton surfaces (see e.g. \cite{Bob1990, Bobbook, Heleinbook2001, Kono1996, KonLand1999, KonoTaim1996, RogSchbook2000, Taim2006} and references therein).  For example, it was shown that integrable dynamics of these surfaces inherit the remarkable properties of integrable equations. In particular, there is a direct connection between certain classes of 2D surfaces and trajectories of infinite dimensional integrable Hamiltonian systems. 

One area of much recent activity in the study of soliton surfaces makes use of the generalized symmetries of the linear spectral problem associated with an integrable equation in order to induce surfaces immersed in  multi-dimensional space. This line of research was begun by A Sym \cite{Sym} and J Tafel \cite{Tafel}  who exploited the symmetry of the linear spectral problem with respect to the spectral parameter to produce a formula for the immersion of surfaces in Lie algebras. The Sym-Tafel formula was extended by adding continuous deformations, including those generated by the gauge symmetry of the linear spectral problem \cite{ Cies1997,FG, Liuthesis}.  Further deformations of the formula were added by A S Fokas, I M Gel'fand F Finkel and Q M Liu \cite{FGFL} who also considered  generalized symmetries of the associated integrable equations and used these symmetries to extend the Sym-Tafel formula for immersion of surfaces in Lie algebras. The immersion function given in \cite{FGFL} has recently been applied by several authors to the study of various classes of soliton surfaces associated with different integrable models (e.g. \cite{  BavMar2009,  BavMar2010, CFG2000}). 

The objective of this paper is to gain a deeper understanding of the formula for immersion given by  Fokas, Gel'fand et al \cite{FGFL} by applying the apparatus of generalized vector fields and their prolongation to this problem. This approach allows us to systematically construct and study the induced surfaces from a Lie group point of view. In particular, we identify a sufficient condition for the generalized symmetries to guarantee that the immersion function can be explicitly integrated. We further propose a formula, analogous to the formula in \cite{FGFL}, for an explicit form of the immersion function and give the necessary and sufficient conditions for the immersion function to be expressible in the proposed form.

The paper is organized as follows. In Section 2, we give a brief overview of the results from  \cite{FGFL}  and in Section 3 we recast these results in terms of generalized symmetries, their characteristic equations, associated vector fields and prolongation structure. In Section 4, we treat separately three symmetries of the linear spectral problem and its compatibility conditions, namely  conformal symmetry in the spectral parameter, a gauge transformation of the wave functions for the linear spectral problem and generalized symmetries of an integrable equation equivalent to the compatibility conditions of the linear spectral problem. In this section, we define the Fokas-Gel'fand immersion function as the immersion function generated by a linear combination of these three symmetries. The surface is determined by its tangent vectors and defined up to a constant of integration chosen in such a way that the immersion function is in the Lie algebra. We further propose an explicit formula for the integration of the immersion function, analogous to the formula defined in \cite{FGFL}, and show that tangent vectors  coincide with those given in \cite{FGFL} if and only if the generalized symmetry of the integrable equation is also a symmetry of the linear spectral problem. The proposed formula is expressed in terms of an arbitrary real function of the spectral parameter, an arbitrary gauge function in the Lie algebra, and the prolongation of the evolutionary form of the generalized vector field acting on wave functions for the linear spectral problem. The formula is given in equation (\ref{fgf}). In Section 5, we illustrate these theoretical considerations by considering the completely integrable two dimensional $\mathbb{C}P^{N-1}$ sigma model and the immersion functions associated with the conformal symmetry of the equations of motion. In particular, we show that the Fokas-Gel'fand formula for immersion can be integrated for the $\mathbb{C}P^{N-1}$ sigma model defined on either Minkowski or Euclidean space. However, in the case of traveling wave solutions of the  $\mathbb{C}P^{N-1}$ sigma model defined on Minkowski space, the Fokas-Gel'fand immersion formula will have the proposed form (\ref{fgf}) if and only if the conformal symmetry is generated by a dilation and translation of the space of independent variables. On the other hand, for finite action solutions of the  $\mathbb{C}P^{N-1}$ sigma model defined on Euclidean space, a general conformal symmetry is also a symmetry of the linear spectral problem and so the Fokas-Gel'fand formula will have the proposed form (\ref{fgf}). 

\section{The Fokas-Gel'fand immersion formula }
In this section, we briefly summarize the results of A S Fokas, I M Gel'fand et al \cite{FGFL} in order to provide a basis for  further analysis in this paper. 
Supppose that the following matrix system of non-linear partial differential equations  (NPDE's) in two independent variables $u,$ and  $v$ of the form 
\be \label{ce}\Delta\equiv  \partial_vU-\partial_uV+[U,V]=0\ee
 admits a  linear spectral problem (LSP) given  by 
\be \label{lsp} \partial_u \Phi =U\Phi, \qquad \partial_v\Phi =V \Phi, \ee
where $\Phi(u,v, \lambda)$ takes its values in some Lie group $G$ and $ U(u,v,\lambda)$ and $ V(u,v,\lambda)$ are matrix functions in the associated Lie algebra $\mathfrak{g}.$
Note that, as long as $U$ and $ V$ satisfy (\ref{ce}) there exists some group-valued function $\Phi$ which satisfies the LSP (\ref{lsp}).

Now, consider an infinitesimal symmetry  of the matrix system of NPDE's (\ref{ce}) and its associated  LSP (\ref{lsp})   given by 
\be\label{sym} \left( \ba{c} U' \\ V'\\ \Phi' \ea \right)= \left( \ba{c} U \\ V \\ \Phi \ea \right)+\epsilon \left( \ba{c} A \\ B \\ \Psi  \ea \right), \qquad 0<\epsilon\in \mathbb{R} \ee
with the requirements that $U',\ V' \in \g,$ and  $\Phi'\in G.$ Such deformations are also required to have the same singularity structure as $U$ and $ V,$ in $\lambda$ given by (\ref{ce}) and (\ref{lsp}). That is,  the matrices $A(u,v, \lambda), \ B(u,v, \lambda)$ taking values in the Lie algebra $\g$ and $\Psi(u,v)$ such that $\Phi+ \epsilon \Psi $ is in the Lie group $G$ will also obey some differential constraints arising from the infinitesimal deformations of (\ref{ce}) and (\ref{lsp}).
The infinitesimal deformation of the  NPDE's, (\ref{ce}) requires that the $g$-valued matrix functions $A,B$ satisfy 
\be \label{celsp} \frac{\partial A}{\partial v}-\frac{\partial B}{\partial u} +[A,V] +[U,B]=0.\ee
and the infinitesimal deformation of (\ref{lsp}) implies
\be \label{tanlsp} \frac{\partial \Psi}{\partial u} =U\Psi +A\Phi, \qquad \frac{\partial \Psi}{\partial v} =V\Psi +B\Phi.\ee 

According to \cite{FG, FGFL}, an infinitesimal symmetry of the integrable equations (\ref{ce}), given by matrix functions $A,B \in \g$ which satisfy (\ref{tanlsp}), suffices to  generate a surface immersed in the Lie algebra $\g$ and also to induce an infinitesimal symmetry of the LSP (\ref{lsp}).  We present these results in the following theorem and corollary.

\begin{theorem}\textbf{Fokas, Gel'fand et al \cite{FGFL}} Suppose that we have matrix functions $U, V\in \g$, and $\Phi\in G$ which satisfy the matrix system of NPDE's (\ref{ce}) and the associated LSP (\ref{lsp}).
Suppose further that $A,B \in \g$ are matrix functions which satisfy (\ref{celsp}). Then, there exists a $\g$-valued immersion function $F(u,v)$ such that the tangent vectors to the 2D surface are given by  
\be \label{cethm} \frac{\partial F}{\partial u}=\Phi^{-1} A \Phi, \qquad \frac{\partial F}{\partial v}=\Phi^{-1} B \Phi.\ee
\end{theorem}
\textbf{Proof.} Equations  (\ref{celsp}) are exactly the compatibility equations for (\ref{cethm}) and so the immersion function $F$ exists and, up to affine transformations, can be assumed to be in the Lie algebra $\g$. \qed 
As a corollary, it was shown  \cite{FG, FGFL} that from this surface one can construct an infinitesimal symmetry of not just the integrable NPDE's (\ref{ce}) but also of its  LSP (\ref{lsp}) .

\begin{corollary}
Suppose that we have matrix functions $U, V\in \g$ and $\Phi\in G$ which satisfy (\ref{ce}) and (\ref{lsp}) and  matrix functions $A,B \in \g$ which satisfy (\ref{celsp}), then there exists a matrix function $\Psi(u,v)$ such that  $A,B, \Psi $ define an infinitesimal symmetry of equations (\ref{ce}) and (\ref{lsp}). That is, the matrix functions  $A,B,$ and $\Psi$ satisfy (\ref{celsp}) and (\ref{tanlsp}). 
\end{corollary}
\textbf{Proof.} We need to show that given $U, \ V, \ A, \ B \in \g$ and $\Phi \in G$ satisfying (\ref{ce}), (\ref{lsp}) and (\ref{celsp}), there exists some matrix function $\Psi$ such that $\Phi +\epsilon \Psi \in G$ and which satisfies (\ref{tanlsp}). We know by Theorem 1, that there exists a $\g$-valued immersion function  $F$ whose tangent vectors satisfy (\ref{cethm}).  If we define
\be\label{psi}\Psi\equiv \Phi F,\ee
 it is straightforward to determine, using (\ref{cethm}),  that the function $\Psi$ satisfies (\ref{tanlsp}). Furthermore, since $F $ is in the Lie algebra $\g$ the formula
\be \Phi'= \Phi +\epsilon \Psi=\Phi(1+\epsilon F)\ee
implies that $\Phi' $ is in the Lie group $G$. 
Thus, we have constructed an appropriate infinitesimal deformation of the wave function $\Phi.$ 
 \qed 

Further, in \cite{FGFL}, the authors consider admissible symmetries of the system of NPDE's (\ref{ce}) including a conformal transformation of the spectral parameter $\lambda,$ gauge transformations of the wave function $\Phi$, and generalized symmetries of a system of integrable NPDE's equivalent to (\ref{ce}). They then use these symmetries to define a $\g$-valued immersion function of a 2D surface as in Theorem 1. 

That is, suppose that it is possible to parametrize $U$ and $V$ in terms of some set of  dependent variables $\theta^n(u, v)$ $n=1,...,m$ so that the system of NPDE's (\ref{ce}) is independent of spectral parameter $\lambda$ and is an integrable differential system.  Let $a(\lambda)$ be an arbitrary real function of the spectral parameter $\lambda$, $S(u,v,\lambda)$ be an arbitrary matrix function taking values in the Lie algebra $\g$ and $\{\phi^n\}$ be a generalized symmetry of the  integrable system of NPDE's equivalent to (\ref{ce}). Let us define
\bea A&=a \frac{\partial U}{\partial \lambda}+\partial_uS +[S,U]+\sum_{n=1}^m\frac{D U}{D\theta^n}\phi^n,\\
 B&=a \frac{\partial V}{\partial \lambda}+\partial_vS +[S,V]+\sum_{n=1}^m\frac{D V}{D\theta^n}\phi^n,
\eea
where $D/D\theta^n$ is the Fr\'echet derivative in the direction of $\theta^n$. The results of  \cite{FGFL} show that matrices $A$ and $B$ satisfy (\ref{celsp}) and so there exists some $\g$-valued immersion function $F$ with tangent vectors given by 
\be \label{fgtan} \frac{\partial}{\partial u} F=\Phi^{-1} A\Phi, \qquad \frac{\partial}{\partial v} F=\Phi^{-1} B\Phi.\ee
We note here that in order for the immersion function to define a surface, the matrices $A$ and $B$ are required to be linearly independent. 

We shall call any immersion function with  tangent vectors (\ref{fgtan}) a Fokas-Gel'fand immersion. In the paper \cite{FGFL}, the authors propose an integrated form of this surface,
\be \label{fprop} F=\Phi^{-1}\left(a \frac{\partial \Phi}{\partial \lambda}+S\Phi+\sum_{n=1}^m\frac{D \Phi}{D\theta^n}\phi^n\right).\ee
We claim that this form of the surface holds if and only if $\{\phi^n\}$ is a generalized symmetry of the LSP (\ref{lsp}) in addition to the integrable system of NPDE's equivalent to (\ref{ce}).

The objective of this paper is to delve deeper into the understanding of this immersion formula defined by tangent vectors (\ref{fgtan}) and the proposed form of its integration (\ref{fprop}). We will make use of the Lie group structure of generalized symmetries and their associated vector fields. 

\section{Generalized Symmetries and the Linear Spectral Problem}
In order to determine the infinitesimal generalized symmetries of the matrix system of NPDE's (\ref{ce}), we make use of the apparatus of vector fields and their prolongations as presented in the book by P J Olver \cite{Olver}. Specifically, we rewrite the formula for immersion functions of 2D surfaces in Lie algebras in terms of the prolongation formalism of vector fields rather than the notation for Fr\'echet derivatives. In this formulation, we are able to give necessary and sufficient conditions for the existence of a $\g$-valued immersion function $F$ based on the invariance criterion for generalized symmetries \cite{Olver}. Furthermore, we are able to formulate sufficient conditions for the surface $F$ to be explicitly integrated according to tangent vectors (\ref{fgtan}) and show that these conditions are necessary for the $\g$-valued immersion function to have the proposed form (\ref{fprop}).

For uniformity of further computation, let us introduce the following notation   
 
\be u=x^{1}, \quad v=x^2, \quad \lambda=x^3,\quad  U=u^{1}(x^i), \quad V=u^{2}(x^i), \ee
and the derivatives of the matrix functions $u^\alpha $ are given by 
\be u^\alpha_J\equiv \frac{\partial ^{|J|}u^\alpha}{\partial x_{j_1}...\partial x_{j_{|J|}}}, \qquad  |J|>0,\quad  \alpha=1,2 \ee
where $J=(j_1,...j_n)$ is a multi-index with $j_k=1,2,3$ and $|J|=n.$
We decompose the matrix functions $u^{1}$ and $u^{2}$ in the basis $e_j$ $j=1,..,s$ for the Lie algebra $\g$ 
\be u^{1}\equiv u^{1j}e_j \qquad u^{2}\equiv u^{2j}e_j\ee
using the convention that repeated indices are summed over. The $u^{\alpha,j}$ are the dependent variables and $x^1, x^2, x^3$ are the independent variables of the matrix system of NPDE's (\ref{ceO}). In terms of these variables, the total derivative operator $D_J$ is defined iteratively  by
\be D_{J,i}=D_iD_J, \qquad  D_i\equiv \frac{\partial}{\partial x^i}+u^{\alpha j}_{J,i}\frac{\partial}{\partial u^{\alpha j}_{J}}, \qquad i=1,2,3.\ee 
According to the notation in \cite{Olver}, we use the following notation for functions depending on the independent variables $x^i$ and dependent variables $u^{\alpha}$ and their derivatives, $u^\alpha_J$ that is,  $f[u]\equiv f(x^{i},  D_J u^{\alpha j}).$ 

In order to perform the symmetry analysis, we require that the wave function  $\Phi$ in the LSP (\ref{lsp}) depend not only on the independent variables $x^i$ but also on the dependent variables and their derivatives  $u^\alpha_J.$ This assumption is consistent with known solutions of the LSP (e.g in  \cite{DHZ1984} and \cite{FadTakbook}). In what follows, we denote $\widetilde{f}(x^i)\equiv f[u]$  when we would like to refer to the value of the  composition of the mapping $f[u]$ with the dependent variables  and their derivatives, $u^\alpha_J$. In this notation, the wave function and its derivatives can be written as 
 \be \fl\qquad \widetilde{\Phi}(x^i)=\Phi[u]=\Phi(x^i, D_J u^{\alpha,j}(x^i)),\qquad  \frac{\partial}{\partial x^r }\widetilde{\Phi}(x^i)=D_r \Phi [u],\quad r=1,2,3\ee
and the LSP (\ref{lsp}) takes the form 
\be \label{lspO} \frac{\partial}{\partial x^\alpha}\widetilde{\Phi}=u^\alpha\widetilde{\Phi}\iff D_\alpha \Phi=u^\alpha \Phi, \quad  \alpha=1,2\ee
with compatibility conditions given by the matrix system of NPDE's 
\be \label{ceO} u^1_2-u^2_1+[u^1,u^2]=0.\ee

Let $c_{kl}^j$ be the structural coefficients of the Lie algebra $[e_k, e_l]=c_{kl}^j e_j.$ Then the matrix system of  NPDE's (\ref{ceO}) written in this basis is given by the following system of coupled NPDE's  
\be \label{cej} \Delta ^j \equiv u^{1j}_2-u^{2j}_1+u^{1k}u^{2l}c_{kl}^j=0, \quad j=1,..,s.\ee
In what follows, we assume that this matrix system of NPDE's (\ref{cej}) is nondegenerate in the sense given in \cite{Olver}. 
 Consider now a generalized symmetry of the system of  NPDE's  (\ref{cej}) given by the generalized vector field
\be \label{vec} \vec{v}=\sum_{i=1}^{3} \xi^{i}[u]\frac{\partial}{\partial x^{i}} +\sum_{\alpha=1}^{2}\sum_{j=1}^{s}\phi_{\alpha}^j[u]\frac{\partial}{\partial u^{\alpha j}}.\ee
The generalized vector field  $\vec{v}$ written in evolutionary form is
\be \vec{v}_{Q}=\sum_{\alpha=1,2}\sum_{j=1}^{s} Q_{\alpha}^j[u] \frac{\partial}{\partial u^{\alpha, j}}, \qquad Q_{\alpha}^j[u]= (\phi_{\alpha}^ j[u]  - \xi^{i}[u]u^{\alpha j}_i), \quad \alpha=1,2 \ee 
with  $ Q[u]=(Q_1[u], Q_2[u])$ the (matrix) characteristic of $\vec{v}$ where the matrices are defined as $ Q_{\alpha}\equiv Q_{\alpha}^j e_j\in \g.$ The vector field $\vec{v}_Q$ generates an infinitesimal symmetry of the dependent variables $ u^{\alpha,j}$ via the transformation
\be (u^{\alpha,j})'=u^{\alpha,j}+\epsilon Q_\alpha^j, \ee
or equivalently in the matrix form
\be \label{22} (u^{\alpha})'=u^{\alpha}+\epsilon Q_\alpha.\ee

The requirement that the vector field $\vec{v}$ be a generalized symmetry of $\Delta^j=0$ is equivalent to the requirement that the associated evolutionary representative $\vec{v}_Q$ be a generalized symmetry. That is,
\be \label{prcej} pr \vec{v}_Q(\Delta^j)=D_2Q_1^j-D_1Q_2^j+(Q_1^ku^{2l}+u^{1k}Q_2^l)c_{kl}^j=0,\ee
whenever the system of NPDE's (\ref{cej}) holds, ie  $\Delta^j=0.$ 
In matrix form, (\ref{prcej})  becomes 
\be \label{prcem} pr \vec{v}_Q(\Delta)=D_2Q_{1}-D_1Q_{2}+[Q_{1},u^{2}]+[u^1, Q_2]=0,\ee
whenever the matrix NPDE (\ref{ceO}) holds, ie  $\Delta=0.$

Note that the requirements that $\vec{v}$ be a generalized symmetry of $\Delta^j=0$ coincides with the requirements that the infinitesimal deformation (\ref{22}) be an infinitesimal matrix symmetry of the initial equation, $\Delta^j=0.$ Thus, we can use this generalized symmetry to define an immersion function $F[u] \in \g,$  in analogy with Theorem 1, with tangent vectors given by 
\be \label{DF}D_1 F=\Phi ^{-1} Q_1[u] \Phi, \qquad D_2F=\Phi^{-1} Q_2[u] \Phi. \ee
The quantities $Q_1$ and $Q_2$ are analogous to matrices $A$ and $B$ respectively. 
The compatibility equations of (\ref{DF}) are satisfied whenever the vector field $\vec{v}_Q$ given by $Q=(Q_1,Q_2)$ is a generalized symmetry of $\Delta=0$ and so the immersion function $F$ exists and, without loss of generality, can be assumed in to be the Lie algebra $\g.$ 

The following proposition recasts Theorem 1 in the language of generalized vector fields.

\begin{proposition}\label{propu}
Let $\g$ be some Lie algebra with basis $e_j$ and  structural coefficients $c_{kl}^j.$
Assume $u^1=u^{1j}e_j, u^2=u^{2j}e_j\in \g $ are matrix functions which satisfy the system of NPDE's 
\be\label{deltaj} \Delta ^j \equiv u^{1j}_2-u^{2j}_1+u^{1k}u^{2l}c_{kl}^j=0, \quad j=1,..., s\ee  
and   $\Phi[u]\in G$ is a solution of the associated LSP 
\be \label{lspu} D_\alpha \Phi =u^\alpha \Phi , \qquad \alpha=1,2.\ee
Let  $\vec{v}$ be a generalized vector field in dependent variables $u^{1j}, u^{2j}$ and independent variables $x^1, x^2, x^3$  given by 
\be \vec{v}=\sum_{i=1}^{3} \xi^{i}[u]\frac{\partial}{\partial x^{i}} +\sum_{\alpha=1}^{2}\sum_{j=1}^{s}\phi_\alpha^ j[u]\frac{\partial}{\partial u^{\alpha j}}\ee
with evolutionary form 
\be \vec{v}_Q=Q_{\alpha}^j[u] \frac{\partial}{\partial u^{\alpha, j}}, \qquad Q_{\alpha}^j[u]= (\phi_{\alpha}^ j[u]  - \xi^{i}[u]u^{\alpha j}_i), \quad \alpha=1,2 \ee 
Then, the following statements are equivalent: 
\begin{enumerate}
\item \label{1} $\vec{v}$ is a generalized symmetry of $\Delta^j=0.$ 

\item\label{2} There exists an immersion function $\widetilde{F}(x^i)=F[u]$ in the Lie algebra $\g$ with tangent vectors  
\be \label{DFu} \frac{\partial \widetilde{F}(x^i)}{\partial x^\alpha}= D_\alpha F[u]=\Phi^{-1} Q_\alpha[u] \Phi, \qquad \alpha=1,2,\ee
where the matrices $Q_\alpha$ are given by $Q_{\alpha}=Q_\alpha^j e_j.$
\item \label{3}There exists a matrix function $\Psi$ such that the infinitesimal deformation 
\be \label{infu} \left( \ba{c} u^{1} \\ u^{2}\\ \Phi \ea \right) \rightarrow \left( \ba{c} u^{1} \\ u^{2}\\ \Phi \ea \right)+\epsilon \left( \ba{c} Q_1[u] \\ Q_2[u] \\  \Psi \ea \right)\ee
gives a generalized infinitesimal symmetry of the system of NPDE's (\ref{deltaj}) together with its LSP (\ref{lspu}).
\end{enumerate}

Furthermore, if statement (\ref{2}) holds then $\Psi=\Phi F$ is an admissible infinitesimal deformation of $\Phi$ and if statement (\ref{3}) holds, then $F=\Phi^{-1}\Psi$ has tangent vectors which coincide with expression (\ref{DFu}).
\end{proposition}

\textbf{Proof.} 
First, we prove that statement (\ref{1}) is equivalent to statement (\ref{2}). From the invariance criterion for generalized symmetries \cite{Olver}, we know that 
$\vec{v}$ is a generalized symmetry of $\Delta^j=0 $ (\ref{deltaj}) if and only if 
\be \label{doublestar}  pr\vec{v}_Q(\Delta^j)=D_2Q_{1}-D_1Q_{2}+[Q_{1},u^{2}]+[u^1, Q_2]=0,\ee
whenever $\Delta^j=0.$ 
On the other hand, an immersion function $F$ with tangent vectors given by (\ref{DFu}) will exist if and only if the compatibility conditions for the tangent vectors (\ref{DFu}) are satisfied.  The compatibility conditions for (\ref{DFu}) are exactly (\ref{doublestar}) and so the immersion function $F$ exists. Since the tangent vectors of $F$ are assumed to be in the Lie algebra $\g$ we can also assume, modulo a constant of integration, that $F$ is in the Lie algebra $\g$ as well. Thus, statement (\ref{1}) is equivalent to statement (\ref{2}). 

Next, we prove that statement (\ref{2}) is equivalent to statement (\ref{3}). As above, statement (\ref{2}) holds if and only if (\ref{doublestar}) holds. The infinitesimal deformation  (\ref{infu}) is a symmetry of the system of  NPDE's (\ref{deltaj}) if and only if conditions (\ref{doublestar}) hold, and so statement (\ref{3}) implies statement (\ref{2}).
Further, the infinitesimal deformation (\ref{infu}) is also a symmetry of the LSP (\ref{lspu}) if and only if conditions (\ref{tanlsp}) hold,  ie 
\be \label{g} D_\alpha(\Psi)=u^\alpha\Psi  +Q_\alpha\Phi, \quad \alpha =1,2.\ee
If we define $\Psi=\Phi F$ as in (\ref{psi}) then $(\Phi)'=\Phi(\mathbb{I}+\epsilon F)$ is in $G$ and 
 (\ref{g}) is satisfied whenever $\Phi$ is a solution of the LSP (\ref{lspu}) and immersion function $F$ has tangent vectors given by (\ref{DFu}). Thus, statement (\ref{2}) implies statement (\ref{3}) and conversely if statement (\ref{2}) holds then 
\be\label{PF} \Psi=\Phi F\ee  is an admissible symmetry for the wave function $\Phi.$

Finally, if statement (\ref{3}) holds then $\Psi$ satisfies conditions  (\ref{tanlsp}) and so the surface defined by $F= \Phi^{-1}\Psi$ has tangent vectors which coincide with (\ref{DFu}).\qed

 Hence, we have shown that a generalized vector field $\vec{v}$  is a symmetry of the system of NPDE's (\ref{ceO}) if and only if there exists a $\g$-valued immersion function $\widetilde{F}(x^i)$ with tangent vectors given by (\ref{DFu}) which in turn is equivalent to the existence  of a generalized infinitesimal symmetry of the system of NPDE's (\ref{deltaj}) together with its LSP (\ref{lspu}). In the next section, we will give explicit examples of such symmetries and construct their associated surfaces immersed in Lie algebras.   

\section{Symmetries of the linear spectral problem}
We can use some known infinitesimal symmetries of  the system of  NPDE's (\ref{ceO}) together with its  LSP (\ref{lspO}) to generate $\g$-valued immersions $\widetilde{F}(x^i)$ of 2D surfaces in analogy with Fokas, Gel'fand et al \cite{FGFL}.  These include a conformal transformation in the spectral parameter $x^3,$ gauge transformations of the wave function $\Phi$, and generalized symmetries of  the system of integrable NPDE's equivalent to (\ref{ceO}). Let us consider each case separately. 
\subsection{\textbf{Conformal symmetry in the spectral parameter, the Sym-Tafel formula}}
The system of  NPDE's (\ref{ceO}) is invariant under a conformal transformation in the spectral parameter $x^3\equiv \lambda.$  The vector field associated with this symmetry has the form 
\be \vec{v}=-a(x^3)\frac{\partial}{\partial x^3}, \qquad \vec{v}_Q=a(x^3)u^{\alpha, j}_3\frac{\partial }{\partial u^{\alpha, j}},\ee
where $a(x^3)$ is an arbitrary real function of the spectral parameter.
This vector field  generates the infinitesimal symmetry 
$(x^3)'= x^3 -\epsilon a(x^3),$ and $ (u^\alpha)'=u^\alpha+\epsilon u^\alpha_3.$
So, by Proposition 1, there exists a $\g$-valued immersion function  $\widetilde{F}(x^i)$ with tangent vectors given by 
\be \label{comp1} \frac{\partial \widetilde{F}^{ST}(x^i)}{\partial x^\alpha}=D_\alpha F^{ST}[u]=a(x^3)\Phi^{-1} u^\alpha_3 \Phi.\ee
Equations (\ref{comp1}) can be integrated to obtain the $\g$-valued immersion function $\widetilde{F}$ for the surface   given by 
\be \tilde{F}^{ST}(x^i)=F^{ST}[u]=a(x^3)\Phi^{-1} \Psi=a(x^3) \Phi^{-1} D_3 \Phi.\ee
This expression is known as the Sym-Tafel formula for immersion \cite{Sym, Tafel}.
As in part 2 of Proposition 1, this also implies that there exists an infinitesimal deformation of both the system of  NPDE's (\ref{deltaj}) and its  LSP (\ref{lspu}) of the form 
\be \left( \ba{c} u^{1} \\ u^{2}\\ \Phi \ea \right) \rightarrow \left( \ba{c} u^{1} \\ u^{2}\\ \Phi \ea \right)+\epsilon a \left( \ba{c} u^1_3 \\  u^2_3 \\   D_3 \Phi \ea \right),\ee
where, as before, we have the matrix function $\Psi$ defined as in (\ref{psi}) by
\be\Psi= \Phi^{-1}F=D_3 \Phi.\ee

\subsection{\textbf{Gauge transformation of the wave function}}
It is possible to perform the analysis in a somewhat ``opposite'' direction by assuming that statement (\ref{3}) in Proposition \ref{propu} is satisfied and using the infinitesimal symmetry to determine the generalized vector field and the $\g$-valued immersion function. That is, the system of  NPDE's (\ref{deltaj}) and its  LSP (\ref{lspu}) are invariant under the gauge transformation $S(x^i) \in \g$
 \be \label{s}\Phi'= \Phi+ \epsilon \Psi, \qquad \Psi= S\Phi \ee 
associated with the  infinitesimal transformation of the system
\be \left( \ba{c} u^{1} \\ u^{2}\\ \Phi \ea \right) \rightarrow \left( \ba{c} u^{1} \\ u^{2}\\ \Phi \ea \right)+\epsilon \left( \ba{c} \partial_1S +[S,u^1]\\ \partial_2 S +[S, u^2] \\  S \Phi \ea \right),\ee
which is a symmetry of  the system of  NPDE's (\ref{deltaj}) together with its  LSP (\ref{lspu}).

The vector field which generates this infinitesimal symmetry written in evolutionary representation is  
\be \vec{v}_Q=\left( \partial_1 S^j +S^ku^{2l}c_{kl} ^j \right)\frac{\partial}{\partial u^{1j} }+\left( \partial_2S^j +S^ku^{1 l}c_{kl}^j\right)\frac{\partial}{\partial u^{2j}}.\ee
and the $\g$-valued immersion function 
\be \tilde{F}^S(x^i)=F^{S}[u]=\Phi^{-1} \Psi= \Phi^{-1} S \Phi\ee
has tangent vectors 
\be \label{comp2} \frac{\partial \widetilde{F}^{S}(x^i)}{\partial x^ \alpha}=D_\alpha F^{S}[u]=\Phi^{-1} (\frac{\partial S}{\partial x^{\alpha}}+[S,u^\alpha]) \Phi,\ee
consistent with the tangent vectors given in (\ref{DFu}).

\subsection{\textbf{Generalized symmetries of an associated integrable equation}}
Finally, we shall show how a generalized symmetry of some system of integrable NPDE's equivalent to (\ref{ceO}) can be used to satisfy the requirements of Proposition \ref{propu} and to induce a 2D surface immersed in the Lie algebra, $\g.$

Suppose that it is possible to reparametrize the matrices $u^{1}(x^1,x^2,x^3), u^{2}(x^1,x^2,x^3)$ by $m$ unknown functions $\theta^n(x^1,x^2),\ n=1,.., m$ and their derivatives so that the system of NPDE's (\ref{ceO}) is equivalent to  some set of differential equations in these unknown functions, denoted  $M^k$, which are independent of the spectral parameter $x^3$: that is, 
\be \label{Mk} \Delta[u^\alpha([\theta],x^3)]=0 \iff M^k[\theta] =0.\ee
We make use of the notation $f[\theta]=f(x^1,x^2, \theta^n_J)$ and $f([\theta],x^3)=f(x^1,x^2,x^3, \theta^n_J).$

The LSP associated to the integrable system of equations (\ref{Mk}) is given in terms of the wave function $\Phi= \Phi([\theta],x^3)$ by 
\be \label{lspt} D_\alpha \Phi([\theta],x^3)=u^\alpha([\theta],x^3)\Phi([\theta],x^3), \qquad \alpha=1,2.\ee
The integrable system of NPDE's (\ref{Mk}) has dependent variables $\theta^n$ and independent variables $x^1,x^2$ and so an associated   generalized vector field is given by  
\be \vec{w}=s^\alpha[\theta]\frac{\partial}{\partial x^\alpha}+\sum _{n}\phi_{n}[\theta]\frac{\partial }{\partial \theta^n}, \qquad \alpha=1,2.\ee
with  evolutionary form  $\vec{w}_\Q$  
\be \vec{w}_\mathcal{\Q}=\sum _{n}\mathcal{\Q}_{n}[\theta]\frac{\partial }{\partial \theta^n}, \qquad\mbox{where } \mathcal{\Q}_{n}[\theta]\equiv \phi_{n}[\theta]-s^\alpha[\theta]\theta_\alpha^n.\ee
Here, we no longer consider the spectral parameter $x^3$ as an independent variable and so the multi-index $J=(j_1,..., j_n)$ takes values $j_k=1,2.$  Similarly, the total derivative in the new variables is given by 
\be  D_\alpha=\frac{\partial}{\partial x^\alpha}+\theta^n_{J,\alpha}\frac{\partial}{\partial \theta^n_J}, \qquad \alpha =1,2.\ee
This generalized vector field induces an infinitesimal deformation of the functions $\theta^n$ by
\[ (\theta^n)'=\theta^n+\epsilon \Q_n[\theta]\]
and hence induces an infinitesimal deformation of the matrices $u^\alpha$ via 
\[ (u^\alpha)'=u^\alpha +\epsilon pr\vec{w}_\Q u^\alpha([\theta],x^3).\]

As we shall prove in the following proposition, making use of Proposition \ref{propu}, a generalized vector field which is a symmetry of the integrable system of NPDE's (\ref{Mk}) can be used to define a $\g$-valued immersion function $F$ and associated infinitesimal deformation of the integrable system of NPDE's (\ref{Mk}) together with its associated LSP (\ref{lspt}). 
 
\begin{proposition} \label{propw}
 Suppose that there exists an integrable system of NPDE's given in terms of dependent variables $\theta^n(x^1, x^2)$, with $u^\alpha\equiv u^\alpha([\theta],x^3)\in \g$, equivalent to 
\be \label{c} \Delta[u^\alpha([\theta],x^3)]=D_2u^1-D_1u^2+[u^1, u^2]=0\ee 
which is independent of spectral parameter $x^3$ and is the compatibility conditions of the LSP
\be \label{d} D_\alpha \Phi=u^\alpha([\theta],x^3) \Phi,\qquad \alpha=1,2 \ee
for matrix function $\Phi=\Phi([\theta],x^3).$ Consider a generalized vector field $\vec{w}$ in evolutionary form 
\be\vec{w}_\mathcal{\Q}=\sum _{n}\Q_{n}[\theta]\frac{\partial }{\partial \theta^n}.\ee

Then, the following  statements are equivalent: 
\begin{enumerate}
\item \label{1w} The vector field $\vec{w}_\Q$ is a generalized symmetry of the integrable system of NPDE's (\ref{c}).

\item\label{2w} There exists an immersion function $\widetilde{F}(x^i)=F([\theta],x^3)$ in the Lie algebra $\g$ with tangent vectors  
\be \label{DFtw} \frac{\partial \widetilde{F}}{\partial x^\alpha}(x^i)= D_\alpha F([\theta],x^3)=\Phi^{-1} pr\vec{w}_\Q u^\alpha \Phi, \qquad \alpha=1,2.\ee

\item \label{3w}There exists a matrix function $\Psi$ so that the infinitesimal deformation 
\be \label{infw} \left( \ba{c} u^{1} \\ u^{2}\\ \Phi \ea \right) \rightarrow \left( \ba{c} u^{1} \\ u^{2}\\ \Phi \ea \right)+\epsilon \left( \ba{c} pr\vec{w}_\Q u^1 \\ pr\vec{w}_\Q u^2\\  \Psi \ea \right)\ee
gives a generalized infinitesimal symmetry of the system of integrable NPDE's (\ref{c}) together with its LSP (\ref{d}).
\end{enumerate}
Furthermore, if statement (\ref{2w}) holds then $\Psi=\Phi F$ is an admissible infinitesimal deformation of $\Phi$ and if statement (\ref{3w}) holds, then the immersion function $F=\Phi^{-1}\Psi$ has tangent vectors consistent with (\ref{DFtw}).
\end{proposition}
\textbf{Proof:} For the proof, we make use of Proposition \ref{propu} to observe that statement (\ref{2w}) is equivalent to statement (\ref{3w}) and so we need only show that statement (\ref{1w}) is equivalent to statement (\ref{2w}). 

From the invariance criterion for generalized symmetries \cite{Olver}, the generalized vector field
$\vec{w}_\Q $ is a generalized symmetry of (\ref{c})  if and only if 
\be \label{pwD} \fl \quad pr\vec{w}_\Q (D_2u^{1})-pr\vec{w}_\Q (D_1u^{2})+[(pr\vec{w}_\Q u^{1}),u^2]+[u^1,(pr\vec{w}_\Q u^{2})]=0.\ee
On the other hand, the compatibility conditions for the tangent vectors (\ref{DFtw}) are 
 \be \label{doublestarw}\fl \quad  D_2(pr\vec{w}_\Q u^{1})-D_1(pr\vec{w}_\Q u^{2})+[(pr\vec{w}_\Q u^{1}),u^2]+[u^1,(pr\vec{w}_\Q u^{2})]=0.\ee
As was shown \cite{Olver} in Lemma 5.12 (see Appendix), the prolongation of a general evolutionary vector field $\vec{w}_\Q $  commutes with the total derivative $D_\alpha$
 and so (\ref{pwD}) is equivalent to (\ref{doublestarw}). Thus, if $\vec{w}_\Q $ is a generalized symmetry of (\ref{c}), then the compatibility equations for (\ref{DFtw}) are satisfied and so the matrix function $F$ exists and, modulo a constant of integration, can be assumed to be in the Lie algebra $\g.$  Conversely, if an immersion function $F$ exists then (\ref{pwD}) is satisfied and so $\vec{w}_\Q $ is a generalized symmetry of (\ref{c}). Thus, we have proved the theorem. \qed

Finally, we formulate some conditions on the generalized vector field $\vec{w}_\Q $ which allow for the integration of the $\g$-valued immersion function $\widetilde{F}(x^i)$. If the generalized vector field is a symmetry of both the system of integrable NPDE's (\ref{c}) and its LSP (\ref{d}), then the immersion function can be explicitly integrated. We present this result in the following proposition.

\begin{proposition}\label{propwlsp}
Suppose that, as in Proposition \ref{propw}, there exists an integrable system of NPDE's given in terms of dependent variables $\theta^n(x^1, x^2)$, with $u^\alpha\equiv u^\alpha([\theta],x^3)\in \g$, equivalent to (\ref{c})
which is independent of spectral parameter $x^3$ and is the compatibility conditions of the LSP (\ref{d}) for matrix function $\Phi=\Phi([\theta],x^3).$
Suppose further that the generalized vector field given in evolutionary form by
\be\vec{w}_\Q =\sum _{n}\Q _{n}[\theta]\frac{\partial }{\partial \theta^n}.\ee
 is a symmetry of the integrable system of NPDE's $(\ref{c}).$ Then, the following statements are equivalent 

\begin{enumerate}
 \item \label{1lsp}The generalized vector field $\vec{w}_\Q $ is a  symmetry of the LSP  $(\ref{d})$ in the sense that 
\be\label{prwlsp} pr\vec{w}_\Q \left(D_\alpha \Phi-u^\alpha\Phi\right)=0 \mbox{ whenever }D_\alpha \Phi-u^\alpha\Phi=0.\ee
\item \label{2lsp} The $\g$-valued immersion function given by $\widetilde{F}(x^i)$ 
\be \label{e} \widetilde{F}(x^i)=F([\theta],x^3)=\Phi^{-1} pr\vec{w}_\Q \Phi\ee
has  tangent vectors 
\be  \label{DFttt} \frac{\partial \widetilde{F}(x^i)}{\partial x^\alpha} =D_\alpha F([\theta],x^3)=\Phi^{-1} pr\vec{w}_\Q u^\alpha \Phi, \qquad \alpha=1,2.\ee
\item\label{3lsp} The infinitesimal deformation 
\be \label{infwt} \left( \ba{c} u^{1} \\ u^{2}\\ \Phi \ea \right) \rightarrow \left( \ba{c} u^{1} \\ u^{2}\\ \Phi \ea \right)+\epsilon \left( \ba{c} pr\vec{w}_\Q u^1 \\ pr\vec{w}_\Q u^2\\  pr\vec{w}_\Q  \Phi \ea \right)\ee
gives a generalized infinitesimal symmetry of the system of integrable NPDE's (\ref{c}) together with its LSP (\ref{d}).
\end{enumerate}

\end{proposition}

\textbf{Proof.} By  Proposition \ref{propu}, statements (\ref{2lsp}) and (\ref{3lsp}) of Proposition \ref{propwlsp} are equivalent. So, to prove this proposition, we need only show that $\widetilde{F}(x^i)$ has the appropriate tangent vectors (\ref{DFttt}) if and only if $\vec{w}_\Q $ is a symmetry of the LSP  $(\ref{d})$. 
Suppose $\Phi$ satisfies the LSP (\ref{d}). Then from (\ref{DFttt}) and making used of (\ref{d}), (\ref{e}) and (\ref{A1}) we get  
\bea \fl D_\alpha F([\theta],x^3)-\Phi^{-1} pr\vec{w}_\Q u^\alpha \Phi&=D_\alpha( \Phi^{-1} pr\vec{w}_\Q \Phi)-\Phi^{-1} pr\vec{w}_\Q (u^\alpha) \Phi\nn
&= \Phi^{-1}\left(-u^\alpha pr\vec{w}_\Q \Phi +D_\alpha(pr\vec{w}_\Q \Phi)\right)-\Phi^{-1} pr\vec{w}_\Q (u^\alpha) \Phi\nn
&=\Phi^{-1}\left(-u^\alpha pr\vec{w}_\Q \Phi +pr\vec{w}_\Q (D_\alpha\Phi)\right)-\Phi^{-1} pr\vec{w}_\Q (u^\alpha) \Phi\nn
&=\Phi^{-1}\left( pr\vec{w}_\Q (D_\alpha\Phi)- pr\vec{w}_\Q (u^\alpha \Phi)\right)\\
\label{62} &=\Phi^{-1}\left(\prw(D_\alpha\Phi-u^\alpha\Phi)\right).\eea
Thus, the proposed $\widetilde{F}(x^i)$ has the appropriate tangent vectors (\ref{DFttt}), i.e. equation (\ref{62}) is equal to zero, if and only if $\prw(D_\alpha\Phi-u^\alpha\Phi)=0$, that is, the generalized vector field  $\vec{w}_\Q $ is a symmetry of the LSP  $(\ref{d})$. Further, the surface $F$ given by (\ref{e}) is in adjoint representation and so is an element of the Lie algebra $\g.$ \qed

We can compare these results with those in \cite{FGFL} where the immersion is given instead in terms of the Fr\'echet derivative. These two formulations are related via Proposition 5.2 in the book of P J Olver \cite{Olver} which relates the prolongation of $\vec{w}_\Q $ to the Fr\'echet derivative via the formula
\[pr\vec{w}_\Q  \Phi=(\frac{D \Phi}{D \theta^n})\Q _n\equiv \lim_{\epsilon \rightarrow 0}\frac{\partial}{\partial \epsilon}\Phi(\theta^n+\epsilon \Q _n). \]
and similarly
\[ u^\alpha\rightarrow u^\alpha+\epsilon (pr\vec{w}_\Q u^\alpha)=u^\alpha +\epsilon \frac{Du^\alpha}{D\theta^n}\Q _n.\]
 Proposition \ref{propwlsp} implies that, for an infinitesimal symmetry of the system of integrable equations (\ref{c}), the immersion function of the surface given by 
\[ F=\Phi^{-1} (\frac{D \Phi}{D \theta^n})\Q _n\]
will have tangent vectors 
\[D_1F=\Phi^{-1} (\frac{D u^1}{D \theta^n})\Q _n\Phi, \qquad D_2F=\Phi^{-1} (\frac{D u^2}{D \theta^n})\Q _n\Phi,\]
if and only if the symmetry is also an infinitesimal symmetry of the LSP (\ref{d}).

\subsection{\textbf{A linear combination of these symmetries}}
To recapitulate the results from this section, we can take a linear combination of each symmetry and obtain the associated surfaces. 

Suppose that there exists dependent variables $\theta^n(x^1, x^2)$ with $u^\alpha\equiv u^\alpha([\theta],x^3)\in\g$ 
such that the integrable system of NPDE's (\ref{c})
is independent of spectral parameter $x^3$ and is equivalent to the solvability of the LSP (\ref{d}) for the matrix function $\Phi= \Phi([\theta],x^3).$ Let $a(x^3)$ be an arbitrary real function of the spectral parameter $x^3$, $S(x^1,x^2,x^3)$ be an arbitrary matrix function with values in the Lie algebra $\g$ and $\vec{w}_\Q $ a generalized vector field in evolutionary form which is a symmetry of the integrable system of NPDE's (\ref{c}).

Let us define
\bea A&=a u^1_3+\partial_1S +[S,u^1]+\prw u^1\in \g,\\
 B&=a u^2_3+\partial_2S +[S,u^2]+\prw u^2\in\g.
\eea
Then the results of this section imply that there exists some $\g$-valued immersion function $F$ with tangent vectors 
\be \label{FGimm} D_1 F=\Phi^{-1} A\Phi, \qquad D_2 F=\Phi^{-1} B\Phi.\ee
We shall call any immersion function with these tangent vectors a Fokas-Gel'fand immersion. The results of the previous subsections 4.1, 4.2 and 4.3 then show that the Fokas-Gel'fand immersion defined by tangent vectors (\ref{FGimm}) can be integrated explicitly  
\be \label{fgf} F=\Phi^{-1}\left(a(x^3)D_3\Phi+S\Phi +\prw\Phi\right)\in \g\ee
if and only if $\vec{w}_\Q $ is also a symmetry of the LSP (\ref{d}).  

Similarly, there exists some matrix function $\Psi$ such that the infinitesimal deformation 
\[\left( \ba{c} u^{1} \\ u^{2}\\ \Phi \ea \right) \rightarrow \left( \ba{c} u^{1} \\ u^{2}\\ \Phi \ea \right)+\epsilon \left( \ba{c} A \\ B\\  \Psi\ea \right)\]
is a symmetry of the integrable system of NPDE's (\ref{c}) together with its LSP (\ref{d}). The matrix function $\Psi$ is given modulo a constant matrix by 
\[\Psi=a(x^3)D_3\Phi+S\Phi +\prw\Phi\]
if and only if $\vec{w}_\Q $ is also a symmetry of the LSP (\ref{d}). Again, note that $\Psi$ and $F$ can be related via (\ref{PF}).

In the following section, we will develop these theoretical considerations using as an example the completely integrable two dimensional $\mathbb{C}P^{N-1}$ sigma model. In particular, we will show that the conformal symmetry of the model induces a Fokas-Gel'fand immersion in $su(N)$ which does not in general coincide with the immersion formula given by (\ref{fgf}).

\section{Example: $\mathbb{C}P^{N-1}$ sigma model and conformal symmetry}
The objective of this section is to consider 2D surfaces immersed in $su(N)$ associated with the completely integrable $\mathbb{C}P^{N-1}$ sigma model in two dimensions. It is well known that this model has an associated LSP of the form above \cite{ZakMik1979, Zakbook}. It has been  computationally expedient to  express the  $\mathbb{C}P^{N-1}$ sigma model in terms of a rank-one Hermitian projector $P(x^1,x^2)$ with $ P^2=P^{\dagger}=P.$ 

Defining matrix functions, 
\be u^1\equiv\frac{2}{1+\lambda}[D_1 P, P], \qquad u^2\equiv\frac{2}{1-\lambda}[D_2P, P]\ee
the Euler-Lagrange (E-L) equations are  
\be \label{EL} \Delta\equiv [D_{12}P, P]=D_2u^1-D_1u^2+[u^1,u^2]=0\ee
which are exactly the compatibility conditions for the LSP of the form  
\be \label{cplsp} D_\alpha\Phi=u^\alpha \Phi, \qquad  \alpha=1,2\ee
where $\lambda=x^3$ is the spectral parameter. 

In the case of the  $\mathbb{C}P^{N-1}$ sigma model defined on Euclidean space, $x^1=\xi, \ x^2=\overline{\xi}$ are complex variables and the matrices $u^\alpha$ satisfy $(u^1)^\dagger= -u^2$ and $\Phi \in SU(N).$ 
For the $\CP$ sigma model defined on Minkowski space, $x^1=x+t,\ x^2=x-t$ are lightcone coordinates and the matrices $u^\alpha$ satisfy $u^1, u^2 \in su(N)$ and $\Phi \in SU(N).$ 
The action of the $\mathbb{C}P^{N-1}$ sigma model is given by \cite{Zakbook, GoldGrund2010}
\be \label{A} A= \int_\Omega tr(P_1P_2)dx^1dx^2, \qquad \Omega \subset \mathbb{R}^2.\ee The E-L (\ref{EL}) equations admit a conformal symmetry given by the vector field 
\be \label{conf} \vec{w}=-f(x^1)\frac{\partial}{\partial x^1}-g(x^2)\frac{\partial}{\partial x^2}\ee
for $f=f(x^1)$ and $g=g(x^2)$ arbitrary complex functions of one variable.

In order to write this vector field $\vec{w}$ in evolutionary form, we would like to write the E-L equations (\ref{EL}) in terms of some set of dependent functions. To accomplish this, it is useful to rewrite the projector $P$ in terms of an element of the Lie algebra $su(N)$ and then to perform the analysis in this basis via rank-$(N-1)$ matrix function 
\be  \theta\equiv i(P-\mathcal{E}) \in su(N) \Rightarrow \quad P=\mathcal{E}-i\theta, \ee

\be \label{theta2} P^2=P\Rightarrow  \theta\cdot \theta=-i\frac{(2-N)}{N}\theta+\frac{(1-N)}{N}\mathcal{E}, \ee
where we introduce the notation $\mathcal{E}=\mathbb{I}/N.$  
In terms of the basis $e_k$ of $su(N),$ the matrix function $\theta$ can be decomposed into $N^2-1$ real functions 
\be\fl \qquad  \theta=\theta^ke_k \in su(N), \quad [e_j, e_l]=c^k_{j,l}e_k, \quad j,k,l=1,..., N^2-1\ee 
with derivatives
\[ \theta_{J}^k=\frac{\partial ^n\theta^k }{\partial j_1..\partial j_n}, \qquad J=(j_1,.., j_n), \ j_i=1,2 \]
as in Section 4.3. The E-L equations (\ref{EL}) then  become
\be \label{ELt} \Delta(\theta)=-[\theta_{12}, \theta]=0, \qquad \Delta^k(\theta)=-\theta_{12}^j\theta^\ell c_{j\ell}^k=0. \ee
In terms of the dependent variables $\theta^k,$ the evolutionary form of vector field $\vec{w}$ is given by  
\be \vec{w}_C\equiv\left(f(x^1)\theta_1^j+g(x^2)\theta_2^j\right)\frac{\partial}{\partial \theta^j}\ee
with prolongation
\be \fl\quad pr \vec{w}_C=\left(f(x^1)\theta^j_1+g(x^2)\theta^j_2\right)\frac{\partial}{\partial \theta^j}+D_J\left(f(x^1)\theta^j_1+g(x^2)\theta^j_2\right)\frac{\partial}{\partial \theta^j_J}.\ee
For the remainder of the paper, fix $\vec{w}_C$ to be the generalized vector field associated with conformal transformations with characteristic
\be C=(C_1,..., C_{N^2-1}), \quad C_j=f(x^1)\theta_1^j+g(x^2)\theta_2^j.\ee
The prolongation of $\vec{w}_C$ acting on the E-L equations (\ref{ELt}) gives
\be \fl \quad pr\vec{w}_C(\Delta^k)= \left(D_{12}\left(f(x^1)\theta^j_1+g(x^2)\theta^j_2\right)\theta^l +\theta^j_{12}\left(f(x^1)\theta^l_1+g(x^2)\theta^l_2\right)\right)c_{jl}^k, \ee
which in matrix form is  
\be \fl \quad  pr\vec{w}_C(\Delta)=[D_{12}\left(f(x^1)\theta_1+g(x^2)\theta_2\right),\theta] +[\theta_{12},\left(f(x^1)\theta_1+g(x^2)\theta_2\right)].\ee
It can be computed directly that $pr\vec{w}_C(\Delta)=0$ whenever $\Delta=0$ for arbitrary complex functions $f$ and $g$ and so the vector field given by (\ref{conf}) is a symmetry of the E-L equations (\ref{ELt}). However, in the case of the $\CP$ sigma model defined on Minkowski space, we require that the infinitesimal deformation of $u^\alpha$ remain in $su(N),$ i.e.  
\be (u^\alpha+\epsilon pr\vec{w}_Cu^\alpha)\in su(N)\ee  which gives the additional constraints that the functions $f$ and $g$ are required to be real.
 On the other hand, for the  $\mathbb{C}P^{N-1}$ model defined on Euclidean space, the requirement is that 
\be (u^1+\epsilon pr\vec{w}_Cu^1)^\dagger=- (u^2+\epsilon pr\vec{w}_Cu^2)\ee
  which gives the restrictions $g(x^2)=\overline{f(x^1)}=\overline{f}(x^2).$  

The conformal symmetry can be used to defined an $su(N)$-valued immersion function via Proposition \ref{propw} and furthermore, the immersion function can be integrated explicitly. We present these results in the following proposition.
\begin{proposition}\label{propconf}
Suppose that the rank-one Hermitian projector $P= \mathcal{E}-i\theta,$ $\theta\in su(N)$ is a solution of the Euler-Lagrange equations for the $\mathbb{C}P^{N-1}$ sigma model and $\Phi([\theta],x^3) \in SU(N)$ is a solution of its associated  LSP.
That is, we define the matrix functions 
\be  \label{ua} u^1=\frac{-2}{1+\lambda}[\theta_1, \theta]\qquad u^2=\frac{-2}{1-\lambda}[\theta_2, \theta].\ee
The E-L equations 
\be \label{z}  \Delta(\theta)\equiv-[\theta_{12}, \theta]=D_2u^1-D_1u^2+[u^1,u^2]=0\ee
 are exactly the compatibility conditions for the LSP given by 
\be \label{x} D_\alpha\Phi =u^\alpha\Phi, \qquad \alpha=1,2.\ee
Let, $\vec{w}_C$ be a generalized vector field associated with a conformal transformation given by 
\be \label{vecw} \vec{w}_C=\left(f(x^1)\theta_1^j+g(x^2)\theta_2^j\right)\frac{\partial}{\partial \theta^j}.\ee
Then the following statements hold. 
\begin{enumerate} 
\item The $su(N)$-valued immersion function $F$ given  by 
\be\label{f} F=\Phi^{-1}( fu^1+gu^2)\Phi \in su(N)\ee
 has tangent vectors
\be \label{h}D_\alpha F=\Phi^{-1}\prwc u^\alpha \Phi.\ee
That is, $F$ is the Fokas-Gel'fand immersion function associated with conformal symmetries of the $\mathbb{C}P^{N-1}$ model. 
\item The  infinitesimal deformation  given by
\be \label{inf1} \left( \ba{c} u^{1} \\ u^{2}\\ \Phi \ea \right) \rightarrow \left( \ba{c} u^{1} \\ u^{2}\\ \Phi \ea \right)+\epsilon \left( \ba{c} pr\vec{w}_Cu^1 \\ pr\vec{w}_Cu^2\\ (fu^1+gu^2)\Phi=\Phi F \ea \right)\ee
 is a symmetry of the E-L equations(\ref{z}) together with its  LSP (\ref{x}).
\end{enumerate}
 \end{proposition}

\textbf{Proof:}
We know from Proposition \ref{propw} that statement (i) of Proposition 4 is equivalent to statement (ii) so we need only show that the $su(N)$-valued immersion function $F$ given by (\ref{f}) has the appropriate tangent vectors to be a Fokas-Gel'fand immersion function associated with conformal symmetries of the $\mathbb{C}P^{N-1}$ sigma model. We compute the action of the generalized vector field $\vec{w}_C$ on the matrices $u^1$ and $u^2$
\bea pr\vec{w}_Cu^1&=\frac{-2}{1+\lambda}\left(f_1[\theta_1, \theta]+f[\theta_{11},\theta]+g[\theta_1, \theta_2]\right)\nn
\label{prwu1}&=D_1(fu^1)+gD_2(u^1),\eea
and
\bea
  pr\vec{w}_Cu^2&=\frac{-2}{1-\lambda}\left(f[\theta_2,\theta_1]+g[\theta_{22},\theta]+g_2[\theta_2,\theta])\right)\nn
 \label{prwu2}&=fD_1(u^2)+D_2(gu^2).\eea

It is then a straightforward computation to verify that $F$ as given by (\ref{f}) has the appropriate tangent vectors. Indeed, carrying out the differentiation in (\ref{h})  gives 
\begin{eqnarray*}  \fl D_1F&=&D_1\left(\Phi^{-1}( fu^1+gu^2)\Phi\right)\nn
\fl&=&\Phi^{-1}\left( -u^1(fu^1+gu^2) +f_1 u^1+fD_1u^1+fu^1u^1+gD_1u^2+gu^2u_1\right)\Phi\nn
\fl&=&\Phi^{-1}\left( D_1(fu^1)+gD_2(u^1)\right) \Phi, \end{eqnarray*}
and 
\begin{eqnarray*} \fl
D_2F&=D_2\left(\Phi^{-1}( fu^1+gu^2)\Phi\right)\nn
\fl &=\Phi^{-1}\left(-u^2(fu^1+gu^2)+fD_2u^1+fu^1u^2+g_2u^2+gD_2u^2+gu^2u^2\right)\Phi\nn
\fl &=\Phi^{-1} \left(fD_1(u^2)+D_2(gu^2)\right) \Phi,\end{eqnarray*}
where  we have used the E-L equations (\ref{z}).
Thus, the immersion function $F$ as given by (\ref{f}) has the appropriate tangent vectors and so is a Fokas-Gel'fand immersion function associated with conformal symmetries of the $\mathbb{C}P^{N-1}$ sigma model (\ref{EL}). Hence we have proved the proposition. 
 \qed

In this section, we have shown that the surfaces induced by conformal symmetries of the $\mathbb{C}P^{N-1}$ sigma model can be explicitly given in terms of arbitrary functions $f(x^1)$ and $ g(x^2),$ the matrices $\Phi, u^\alpha$ and their derivative. In the following sections, we will compare the immersion functions given by (\ref{f}) with those described in Proposition \ref{propwlsp}. In particular, we shall show that these functions do not coincide in the case of traveling wave solutions for the  $\mathbb{C}P^{N-1}$ sigma model defined on Minkowski space but do for finite action (\ref{A}) solutions of the  $\mathbb{C}P^{N-1}$ sigma model defined on Euclidean space.    

\subsection{\textbf{$\CP$ sigma model defined on Minkowski space}}
In the following section, we would like to investigate cases where  the $su(N)$-valued immersion function given by 
\be \label{calf} \mathcal{F}=\Phi^\dagger \prwc \Phi,\ee
is the Fokas-Gel'fand immersion function generated by the generalized vector field  $\vec{w}_C.$ 
That is, we derive conditions on the generalized vector field $\vec{w}_C$ so that the immersion function $\mathcal{F}$ given by (\ref{calf}) has the following tangent vectors
\be D_\alpha \mathcal{F}=\Phi^\dagger \prwc u^\alpha \Phi, \quad \alpha=1,2.\ee
As above, the statement is equivalent to the requirement that the infinitesimal deformation given by 
\[\left( \ba{c} u^{1} \\ u^{2}\\ \Phi \ea \right) \rightarrow \left( \ba{c} u^{1} \\ u^{2}\\ \Phi \ea \right)+\epsilon \left( \ba{c} pr\vec{w}_Cu^1 \\ pr\vec{w}_Cu^2\\ \prwc \Phi \ea \right)\]  be a symmetry of E-L equations (\ref{z}) together with its  LSP (\ref{x}). That is, we consider the cases where $\mathcal{F}$ and $F$ (as in (\ref{f})) differ by at most a constant matrix of integration. In order to compute $\mathcal{F}$ we require an explicit solution of the wave function and so we consider the simplest case: a traveling wave. In this case, we shall show that these results do not hold in general but only for a restricted class of conformal transformations, namely for translations and dilations.

Let  $\theta$ be a traveling wave solution of the E-L equations (\ref{z}) 
\be \label{trv} \theta\equiv \theta(x^1+\kappa x^2), \qquad \kappa\theta_1- \theta_2=0, \kappa\in \mathbb{R}\ee
with the following differential consequences
 \be \label{travconsq} [\theta_{1},\theta_2]=\kappa[\theta_{1},\theta_1]=0, \quad 0=[\theta_{12}, \theta]=\kappa[ \theta_{11}, \theta] =\frac1\kappa[ \theta_{22},\theta].\ee
 Note that (\ref{travconsq}) implies $D_\alpha[\theta_\beta, \theta]=0$ for $\alpha, \beta =1,2$ and so  $u^1$ and $u^2$ are both constant matrices and thus correspond to vacuum solutions \cite{ZakMik1979, HSS1984}. 
 In this case, we can solve for the wave functions which are given by 
 \be \label{travelwave}\Phi=exp\left(2\chi [\theta_1,\theta]\right)(2i\theta-(2-N)\mathcal{E}), \qquad \chi \equiv (\frac{\lambda x^1}{1+\lambda}-\frac{\kappa\lambda x^2}{ (1-\lambda)}).\ee
By straightforward computation, one obtains  that $\Phi$ given by (\ref{travelwave}) satisfies the LSP (\ref{x}) whenever $\theta\equiv \theta(x^1+\kappa x^2).$ We make use of the algebraic restrictions on the matrix function $\theta$ 
\be \label{y}\theta \cdot \theta=\frac{-i(2-N)}N\theta+\frac{1-N}{N}\mathcal{E}\ee
and the differential consequences of (\ref{y})
\be \frac{-i(2-N)}{N}\theta_1=\theta\theta_1+\theta_1\theta, \qquad \theta\theta_1\theta=\frac{N-1}{N^2}\theta_1,\ee
which in turn implies 
 \bea \label{v} \fl[\theta_1,\theta](2i\theta-(2-N)\mathcal{E})&=&2i\theta_1\theta\theta-2i\theta\theta_1\theta-\frac{(2-N)}{N}(\theta_1\theta-\theta\theta_1)\nn
\fl&=&\frac{(2-N)}{N}(\theta_1\theta+\theta\theta_1)+\frac{4i(1-N)}{N^2}\theta_1=-i\theta_1.\eea
As a consequence of (\ref{y}) and (\ref{v}), the total derivative of $\Phi$ is given by 
\bea \fl D_1\Phi&= D_1(exp\left(2\chi [\theta_1,\theta]\right)(2i\theta-(2-N)\mathcal{E})\nn
\fl &=D_1\left(\sum_{a=0}^\infty \frac{(2\chi)^a}{a!}([\theta_1,\theta])^a(2i\theta-(2-N)\mathcal{E})\right)\nn
\fl &= \sum_{\alpha=1}^{\infty}\frac{(2\chi)^{a-1}}{(a-1)!}2D_1(\chi)([\theta_1,\theta])^a(2i\theta-(2-N)\mathcal{E})+\sum_{a=0}^\infty \frac{(2\chi)^a}{a!}([\theta_1,\theta])^a(2i\theta_1)\nn
\fl &+\sum_{\alpha=1}^{\infty}\frac{(2\chi)^{a}}{(a)!}D_1([\theta_1,\theta]^a)(2i\theta-(2-N)\mathcal{E})\nn
\fl &=\sum_{\alpha=1}^{\infty}\frac{(2\chi)^{a-1}}{(a-1)!}2\frac{\lambda}{1+\lambda}([\theta_1,\theta])^a(2i\theta-(2-N)\mathcal{E})-\sum_{a=0}^\infty \frac{(2\chi)^a}{a!}([\theta_1,\theta])^a2[\theta_1,\theta](2i\theta-(2-N)\mathcal{E})\nn
\fl &+\sum_{\alpha=1}^{\infty}\frac{(2\chi)^{a}}{(a)!}D_1([\theta_1,\theta]^a)(2i\theta-(2-N)\mathcal{E})\nn
\label{D1p}\fl &=\left(\frac{-2}{1+\lambda}[\theta_1,\theta]\right)\Phi+\sum_{\alpha=1}^{\infty}\frac{(2\chi)^{a}}{(a)!}D_1([\theta_1,\theta]^a)(2i\theta-(2-N)\mathcal{E})\\
\label{D1}\fl &=\left(\frac{-2}{1+\lambda}[\theta_1,\theta]\right)\Phi \quad \mbox{ whenever }\theta=\theta(x^1+\kappa x^2) \mbox{ and } [\theta_{12},\theta]=0
\eea
In the expression (\ref{D1}), we have used the fact that if $\theta$ is a traveling wave solution of the E-L equations (\ref{z}) then $D_\alpha [\theta_\beta,\theta]=0$ for $\alpha, \beta =1,2.$  Similarly, we compute the derivative of $\Phi$ with respect to $x^2$ to obtain
\bea \fl
D_2\Phi&= D_2(exp\left(2\chi [\theta_1,\theta]\right)(2i\theta-(2-N)\mathcal{E})\nn
\fl &=D_2\left(\sum_{a=0}^\infty \frac{(2\chi)^a}{a!}([\theta_1,\theta])^a(2i\theta-(2-N)\mathcal{E})\right)\nn
\fl &= \sum_{\alpha=1}^{\infty}\frac{(2\chi)^{a-1}}{(a-1)!}2D_2(\chi)([\theta_1,\theta])^a(2i\theta-(2-N)\mathcal{E})+\sum_{a=0}^\infty \frac{(2\chi)^a}{a!}([\theta_1,\theta])^a(2i\theta_2)\nn
\fl &+\sum_{\alpha=1}^{\infty}\frac{(2\chi)^{a}}{(a)!}D_2([\theta_1,\theta]^a)(2i\theta-(2-N)\mathcal{E})\nn
\fl &=\sum_{\alpha=1}^{\infty}\frac{(2\chi)^{a-1}}{(a-1)!}\frac{-2\kappa \lambda}{1-\lambda}([\theta_1,\theta])^a(2i\theta-(2-N)\mathcal{E})\nn
\fl &-\sum_{a=0}^\infty \frac{(2\chi)^a}{a!}([\theta_1,\theta])^a2[\theta_2,\theta](2i\theta-(2-N)\mathcal{E})+\sum_{\alpha=1}^{\infty}\frac{(2\chi)^{a}}{(a)!}D_2([\theta_1,\theta]^a)(2i\theta-(2-N)\mathcal{E})\nn
\label{D2p}\fl &=\left(\frac{-2\kappa\lambda}{1-\lambda}[\theta_1,\theta]-2[\theta_2,\theta]\right)\Phi+\sum_{\alpha=1}^{\infty}\frac{(2\chi)^{a}}{(a)!}D_2([\theta_1,\theta]^a)(2i\theta-(2-N)\mathcal{E})\\
\label{D2}\fl &=\left(\frac{-2}{1-\lambda}[\theta_2,\theta]\right)\Phi\quad  \mbox{ whenever }\theta=\theta(x^1+\kappa x^2) \mbox{ and } [\theta_{12},\theta]=0.
\eea
Again,  we have used  $D_\alpha [\theta_\beta,\theta]=0$ for $\alpha, \beta =1,2$ and $\kappa \theta_1= \theta_2$ whenever $\theta(x^1+\kappa x^2)$ is a solution of the E-L equations (\ref{z}).
Thus,  the wave function $\Phi$ given by (\ref{travelwave}) satisfies the LSP (\ref{x}) whenever $\theta(x^1+\kappa x^2)$ is a solution of the E-L equations (\ref{z}).

In order to present an explicit formula for the $su(N)$-valued immersion function $\mathcal{F},$ we compute the action of the prolongation of the generalized vector field $\vec{w}_C$ on $\Phi$ of the form (\ref{travelwave}). 
The prolongation of the generalized vector field $\vec{w}_C$ simplifies to  
\bea\fl pr \vec{w}_C&=\sum_{J}D_J (f\theta^k_{1}+g\theta^k_{2}) \frac{\partial}{\partial \theta^k_{J}}\nn
 \fl &=f(D_1-\frac{\partial}{\partial x^1})+g(D_2-\frac{\partial}{\partial x^2})+\sum_{n>0,J}(f_{n}\theta^k_{J,1}+g_{n}\theta^k_{J,2}) \frac{\partial}{\partial \theta^k_{n,J}}. 
\eea
Its action on the wave function $\Phi$ is given by 
\bea\label{f1} pr\vec{w}_C\Phi&=f(D_1-\frac{\partial}{\partial x^1})\Phi+f_1\theta^j_1\frac{\partial \Phi}{\partial \theta^j_1}+g(D_2-\frac{\partial}{\partial x^2})\Phi.
\eea
Computing each term separately, we obtain
\bea 
 \label{102} \frac{\partial}{\partial x^\alpha}\Phi&=\frac{\partial}{\partial x^1}\left(\sum_{a=0}^\infty \frac{(2\chi)^a}{a!}([\theta_1,\theta])^a(2i\theta-(2-N)\mathcal{E})\right)=2\frac{\partial\chi}{\partial x^\alpha}[\theta_1,\theta]\Phi.
\eea
and 
\bea \theta^j_1\frac{\partial \Phi}{\partial \theta^j_1}&= \theta^k_1\frac{\partial}{\partial \theta^k_1}exp\left(2\chi \theta_1^n\theta^lc_{nl}^j e_j\right)(2i\theta^me_m-(2-N)\mathcal{E})\nn
\label{105} &= \theta^k_1\frac{\partial}{\partial \theta^k_1}\sum_{a=0}^\infty\frac1 {a!}\left(2\chi \theta_1^n\theta^lc_{nl}^j e_j\right)^a(2i\theta^me_m-(2-N)\mathcal{E}).
\eea
Note that in the above sum, the only non-zero terms are those for which $k=n$ and (\ref{105}) simplifies
\bea \theta^k_1\frac{\partial \Phi}{\partial \theta^k_1}&=\sum_{a=0}^\infty\frac {a(2\chi)^a} {a!}\left( \theta_1^n\theta^lc_{nl}^j e_j\right)^{a}(2i\theta^me_m-(2-N)\mathcal{E})\nonumber\\
&=(2\chi\theta_1^n\theta^lc_{nl}^j e_j)\sum_{a=1}^\infty\frac {(2\chi)^{a-1}} {(a-1)!}\left( \theta_1^n\theta^lc_{nl}^j e_j\right)^{a-1}(2i\theta^me_m-(2-N)\mathcal{E})\nonumber\\
&=(2\chi\theta_1^n\theta^lc_{nl}^j e_j)\Phi\nn
\label{104}&=2\chi [\theta_1,\theta]\Phi.\eea
We then substitute (\ref{102}-\ref{104}) into (\ref{f1})  using the requirement $k\theta_1- \theta_2=0 $ and the fact the $\Phi$ satisfies the LSP (\ref{x}) to obtain a closed expression for the $su(N)$-valued immersion function $\mathcal{F}$
\bea\fl \mathcal{F}=\Phi^\dagger \left(f(D_1\Phi-\frac{2\lambda}{1+\lambda}[\theta_1, \theta]\Phi)+g(D_2\Phi+\frac{\lambda\kappa}{1-\lambda}[\theta_1,\theta])+2f_1\chi[\theta_1,\theta]\right)\Phi\nn
\label{107}=  \left(-2f-2\kappa g+2f_1\chi\right)\Phi^\dagger[\theta_1,\theta]\Phi.\eea
Next, we compute the tangent vectors to the immersion function $\mathcal{F}.$ First,  note that the matrix $\Phi^{-1}[\theta_1,\theta]\Phi$ is constant whenever $\Phi$ satisfies the LSP (\ref{x}) and $\theta$ is a traveling wave with $\theta_2=\kappa \theta_1.$ Computing the derivatives of the matrix $\Phi^{-1}[\theta_1,\theta]\Phi$ gives
\[D_1\left(\Phi^{-1}[\theta_1,\theta]\Phi\right)=\Phi^\dagger\left( -[\theta_1,\theta][\theta_1,\theta]+[\theta_1,\theta][\theta_1,\theta]\right)\Phi
=0,\]
and 
\[D_2\left(\Phi^{-1}[\theta_1,\theta]\Phi\right)=\Phi^\dagger\left( -[\theta_2,\theta][\theta_1,\theta]+[\theta_1,\theta][\theta_2,\theta]\right)\Phi
=0.\]
This greatly simplifies the computation of the tangent vectors and in particular, we can observe that the immersion function is degenerate and gives a curve instead of a 2D surface. We note here that we are only considering the third part of the Fokas-Gel'fand immersion formula defined by tangent vectors (\ref{FGimm}) and in particular if we consider the case where $a(x^3)=0$ and $S(x^i)\in su(N)$ is an arbitrary gauge, the function  
\be F= \left(-2f-2\kappa g+2f_1\chi\right)\Phi^\dagger[\theta_1,\theta]\Phi +\Phi^\dagger S(x^i)\Phi\ee
defines a 2D surface immersed in $su(N).$
 
The tangent vectors for the $su(N)$-valued immersion function $\mathcal{F}$ given by (\ref{107}) are 
\bea \label{d1f} D_1\mathcal{F}&= \left(\frac{-2}{1+\lambda}f_1+f_{11}\chi\right)\Phi^\dagger[\theta_1,\theta]\Phi,\\
\label{d2f} D_2\mathcal{F}&=\left(-2g_2-f_1\frac{2\lambda\kappa}{1-\lambda}\right)\Phi^\dagger[\theta_1,\theta]\Phi .\eea
With such an immersion function, we can then use Proposition \ref{propu} to induce an infinitesimal symmetry of the E-L equations (\ref{z}) together with its LSP (\ref{x}). We present these results in the following proposition. 
\begin{proposition}\label{travg}
Suppose that the rank-one Hermitian projector $P= \mathcal{E}-i\theta,$ $\theta(x^1+\kappa x^2)\in su(N)$ is a a traveling wave solution of the Euler-Lagrange equations (\ref{z}) for the $\mathbb{C}P^{N-1}$ sigma model defined on Minkowski space.
Furthermore,  let $\vec{w}_C$ be the generalized vector field associated with conformal transformations (\ref{vecw}) and $u^\alpha$ defined as in (\ref{ua}). 
Then the following statements hold:
\begin{enumerate} 
\item The $SU(N)$ matrix function $\Phi$ given by 
\be \label{travelwavep}\fl \quad \Phi=exp\left(2\chi [\theta_1,\theta]\right)(2i\theta-(2-N)\mathcal{E}), \qquad \chi \equiv (\frac{\lambda x^1}{1+\lambda}-\frac{\kappa\lambda x^2}{ (1-\lambda)})\ee satisfies the LSP (\ref{x}) 
\item The $su(N)$-valued immersion function $\mathcal{F}$ given  by 
\be\label{fp} \mathcal{F}=\Phi^{-1}pr\vec{w}_C\Phi\ee
 has tangent vectors
\be \label{hp}D_\alpha \mathcal{F}=\Phi^{-1}R_\alpha\Phi, \quad \alpha=1,2\ee
where
\be \fl R_1=\left(\frac{-2}{1+\lambda}f_1+f_{11}\chi\right)[\theta_1,\theta], \quad  R_2=\left(-2g_2-f_1\frac{2\lambda\kappa}{1-\lambda}\right)[\theta_1,\theta].\ee

\item The  infinitesimal deformation  given by
\be \label{infp} \left( \ba{c} u^{1} \\ u^{2}\\ \Phi \ea \right) \rightarrow \left( \ba{c} u^{1} \\ u^{2}\\ \Phi \ea \right)+\epsilon \left( \ba{c} R_1 \\ R_2\\ \Psi= pr\vec{w}_C\Phi \ea \right)\ee
 is a symmetry of the E-L equations(\ref{z}) together with its  LSP (\ref{x}). 
\end{enumerate}\end{proposition}
\textbf{Proof:} We have already proved the results above. First, the computation of $D_\alpha \Phi$  given by (\ref{D1}) and (\ref{D2}) shows that $\Phi$ given by (\ref{travelwavep}) is a solution of the LSP (\ref{x}) whenever $\Phi$ is a traveling wave solution of the E-L equations (\ref{z}). Also, the computations of the tangent vectors $D_\alpha \mathcal{F}$ given by  (\ref{d1f}) and (\ref{d2f}) show that the $su(N)$-valued immersion function $\mathcal{F}$ has tangent vectors which coincide with expression (\ref{hp}) whenever $\Phi$ is a traveling wave solution of the E-L equations (\ref{z}). Finally, we check that the infinitesimal deformation given by (\ref{infp}) is a symmetry of the E-L equations (\ref{z}) together with its  LSP (\ref{x}) modulo the requirement that $\kappa\theta_1-\theta_2=0.$ 
But as before, the requirement that the tangent vectors given by (\ref{hp}) be compatible is exactly given by the requirements that (\ref{infp}) be an infinitesimal symmetry of the E-L equations (\ref{z})
\[D_2R_1-D_1R_2+[R_1,u^2]+[u^1,R_2]=0.\] 
Similarly, we can use the equation for the tangent vectors (\ref{hp}) to show that 
\[D_\alpha (pr\vec{w}_C\Phi)=u^\alpha(pr\vec{w}_C\Phi) +R_\alpha\Phi,\]
whenever $\Phi,$ given by (\ref{travelwavep}), is a solution of the LSP (\ref{x}) and so the infinitesimal deformation given by (\ref{infp}) is  a symmetry of the E-L equations (\ref{z}) together with its  LSP (\ref{x}). \qed
Note that for arbitrary real functions $f$ and $g$ the matrix functions  \[ R_\alpha \ne pr\vec{w}_Cu^\alpha,\quad \alpha=1,2\] and so the tangent vectors to $\mathcal{F}$ are not, in general, given by $D_\alpha \mathcal{F}=\Phi^\dagger \prwc u^\alpha\Phi.$ Thus $\mathcal{F}$ does not generally have the form of a Fokas-Gel'fand immersion function generated by a conformal symmetry (\ref{vecw}) of the $\mathbb{C}P^{N-1}$ sigma model defined on Minkowski space. 
However, we can use Proposition \ref{propwlsp} to give necessary and sufficient conditions on the generalized vector field $\vec{w}_C$ so that the immersion function $\mathcal{F}$ will have tangent vectors given by  $D_\alpha \mathcal{F}=\Phi^\dagger \prwc u^\alpha \Phi$ and so will be the Fokas-Gel'fand immersion function generated by conformal symmetries of the $\mathbb{C}P^{N-1}$ sigma model defined on Minkowski space. We state these results in the following proposition. 

\begin{proposition} \label{proptwlsp}
Suppose that the rank-one Hermitian projector $P= \mathcal{E}-i\theta,$ $\theta(x^1+\kappa x^2)\in su(N)$ is a a traveling wave solution of the Euler-Lagrange equations (\ref{z}) for the $\mathbb{C}P^{N-1}$ sigma model defined on Minkowski space.
Furthermore,  as above, let $\vec{w}_C$ be the generalized vector field associated with conformal transformations of the form (\ref{vecw}) and $u^\alpha$ are defined as in (\ref{ua}). 

Then the  following statements are equivalent:
\begin{enumerate} 
\item \label{1twlsp} The generalized vector field $\vec{w}_C$ is a generalized symmetry of the traveling wave equations (\ref{trv}). That is, $\prwc (\kappa \theta_1-\theta_2)=0$ whenever $\kappa \theta_1-\theta_2=0$.
\item \label{2twlsp} The $su(N)$-valued immersion function $\mathcal{F}$ given  by 
\be\label{fptvlsp} \mathcal{F}=\Phi^{-1}pr\vec{w}_C\Phi\ee
 has tangent vectors
\be \label{dftvlsp}D_\alpha \mathcal{F}=\Phi^{-1}\prwc u^\alpha\Phi ,\quad \alpha=1,2\ee
and so $\mathcal{F}$ is  the Fokas-Gel'fand immersion function generated by conformal symmetries of the $\mathbb{C}P^{N-1}$ sigma model defined on Minkowski space.
\item\label{3twlsp} The  infinitesimal deformation  given by
\be \label{inftvlsp} \left( \ba{c} u^{1} \\ u^{2}\\ \Phi \ea \right) \rightarrow \left( \ba{c} u^{1} \\ u^{2}\\ \Phi \ea \right)+\epsilon \left( \ba{c} pr\vec{w}_Cu^1 \\ pr\vec{w}_Cu^2\\  pr\vec{w}_C \Phi \ea \right).\ee
is a symmetry of the E-L equations(\ref{z}) together with its  LSP (\ref{x}). 
\end{enumerate}
\end{proposition}
\textbf{Proof:}
Using proposition \ref{propwlsp}, all that needs to be shown is that for  $\Phi$  a solution of the LSP (\ref{x}), the condition
\be \label{ref14} pr\vec{w}_C\left(D_\alpha \Phi-u^\alpha\Phi\right)=0 \iff f=ax^1+b,\quad g=ax^2+c\ee
holds.
We have already computed the expression $D_\alpha \Phi-u^\alpha\Phi$ given by  (\ref{D1p}) and (\ref{D2p}) so we need only 
 compute the action of the prolongation of $\vec{w}_C$ on this sum, taken modulo $\theta$ a traveling wave solution of the E-L equations (\ref{z})
\bea \fl pr\vec{w}_C\left(D_1 \Phi-u^1\Phi\right)=\\
 \bigg(f(D_1-\frac{\partial}{\partial x^1})+f_1(\theta^k_1\frac{\partial }{\partial \theta^k_1}+\theta^k_{11}\frac{\partial }{\partial \theta^k_{11}}+\theta^k_{12}\frac{\partial }{\partial \theta^k_{12}})+f_{11}(\theta^k_{1}\frac{\partial }{\partial \theta^k_{11}})\nn
+g(D_2-\frac{\partial}{\partial x^2})+g_2(\theta^k_2\frac{\partial }{\partial \theta^k_2}+\theta^k_{12}\frac{\partial }{\partial \theta^k_{12}}+\theta^k_{22}\frac{\partial }{\partial \theta^k_{22}})\bigg)\left(D_1 \Phi-u^1\Phi\right)\nonumber.
\eea
Invoking (\ref{D1p}), the quantity $D_1\Phi-u^1\Phi$ is given by 
\be D_1\Phi-u^1\Phi= \sum_{\alpha=1}^{\infty}\frac{(2\chi)^{a}}{a!}D_1([\theta_1,\theta]^a)(2i\theta-(2-N)\mathcal{E})\ee
and so we obtain 
\bea \fl \prwc(D_1\Phi-u^1\Phi)=\prwc\left(\sum_{\alpha=1}^{\infty}\frac{(2\chi)^{a}}{a!}D_1([\theta_1,\theta]^a)(2i\theta-(2-N)\mathcal{E})\right)\nn
\label{ref2}\fl\quad =2\prwc(\chi)\sum_{\alpha=1}^{\infty}\frac{(2\chi)^{a-1}}{(a-1)!}D_1([\theta_1,\theta]^a)(2i\theta-(2-N)\mathcal{E}))\\\
+\sum_{\alpha=1}^{\infty}\frac{(2\chi)^{a}}{a!}\prwc\left(D_1([\theta_1,\theta]^a)\right)(2i\theta-(2-N)\mathcal{E})\nn
 +\sum_{\alpha=1}^{\infty}\frac{(2\chi)^{a}}{a!}D_1([\theta_1,\theta]^a)\prwc(2i\theta-(2-N)\mathcal{E})\nonumber.
\eea
Recall, that if $\theta(x^1+\kappa x^2)$ is a traveling wave solution of the E-L equations (\ref{x}), the matrix $[\theta_1, \theta]$ is a constant and hence $D_1[\theta_1,\theta]=0$. Thus, modulo the requirements that $\theta$ be a traveling wave solution of the E-L equations (\ref{x}), the quantity (\ref{ref2}) becomes
\bea \fl\prwc(D_1\Phi-u^1\Phi)&=\sum_{\alpha=1}^{\infty}\frac{(2\chi)^{a}}{a!}\prwc\left(D_1([\theta_1,\theta]^a)\right)(2i\theta-(2-N)\mathcal{E})\nn
\fl\label{ref5}&=\sum_{\alpha=1}^{\infty}\frac{(2\chi)^{a}}{a!}D_1\left(\prwc([\theta_1,\theta]^a)\right)(2i\theta-(2-N)\mathcal{E}).
\eea
The action of $\prwc$ on $[\theta_1,\theta]^a$ is given by
\bea \fl \prwc([\theta_1,\theta]^a)&=(fD_1+gD_2)[\theta_1,\theta]^a&+f_1\left(\theta_1^k\frac{\partial}{\partial \theta_1^k}(\theta_1^n\theta^lc_{nl}^j e_j)^a\right)\nn
 &=af_1(\theta_1^n\theta^lc_{nl}^j e_j)^a &\mbox{ modulo } \kappa\theta_1-\theta_2=0, \quad [\theta_{12},\theta]=0,\nn
\label{ref4}&=af_{1}[\theta_1,\theta]^a &\mbox{ modulo } \kappa\theta_1-\theta_2=0, \quad [\theta_{12},\theta]=0.\eea
Thus, 
\[D_1\prwc([\theta_1,\theta]^a)=af_{11}[\theta_1,\theta]^a \mbox{ modulo } \kappa\theta_1-\theta_2=0, \quad [\theta_{12},\theta]=0,\]
and  (\ref{ref5}) simplifies to 
\be prw(D_1\Phi-u^1\Phi)=\sum_{\alpha=1}^{\infty}\frac{(2\chi)^{a}}{(a-1)!}f_{11}[\theta_1,\theta]^a(2i\theta-(2-N)\mathcal{E})\ee
which is equal, via the  LSP (\ref{x}), to
\be pr\vec{w}_C \left(D_1 \Phi-u^1\Phi\right)=-f_{11}\chi(1+\lambda)D_1\Phi.\ee
Hence, for non-constant wave functions, $pr\vec{w}_C \left(D_1 \Phi-u^1\Phi\right)$ vanishes  if and only if $f_{11}=0, $ ie $f=ax^1+b.$

For $\alpha=2$, expression (\ref{D2p}) gives
\be \fl D_2\Phi-u^2\Phi= \left(\frac{-2\kappa\lambda}{1-\lambda}[\kappa\theta_1-\theta_2,\theta]\right)\Phi+\sum_{\alpha=1}^{\infty}\frac{(2\chi)^{a}}{(a)!}D_2([\theta_1,\theta]^a)(2i\theta-(2-N)\mathcal{E}).\ee
The computation of the action of $\prwc$ on $D_2\Phi-u^2\Phi$  is similar to those above, 
\bea\fl \prwc(D_2\Phi-u^2\Phi)=\prwc\left[\left(\frac{-2\kappa\lambda}{1-\lambda}[\kappa\theta_1-\theta_2,\theta]\right)\right]\Phi +\left(\frac{-2\kappa\lambda}{1-\lambda}[\kappa\theta_1-\theta_2,\theta]\right)\prwc\Phi\nn
\label{ref7}\fl \quad +2\prwc(\chi)\sum_{\alpha=1}^{\infty}\frac{(2\chi)^{a-1}}{(a-1)!}D_2([\theta_1,\theta]^a)(2i\theta-(2-N)\mathcal{E}))\\
 +\sum_{\alpha=1}^{\infty}\frac{(2\chi)^{a}}{a!}\prwc\left(D_2([\theta_1,\theta]^a)\right)(2i\theta-(2-N)\mathcal{E})\nn
+\sum_{\alpha=1}^{\infty}\frac{(2\chi)^{a}}{a!}D_2([\theta_1,\theta]^a)\prwc(2i\theta-(2-N)\mathcal{E})\nonumber.
\eea
Again, if $\theta$ is a traveling wave solution of the E-L equations (\ref{x}), then $\kappa\theta_1-\theta_2=0$ and $D_2[\theta_1,\theta]=0,$ and so the quantity (\ref{ref7}) becomes
\bea\fl \prwc(D_2\Phi-u^2\Phi)=\prwc\left[\left(\frac{-2\kappa\lambda}{1-\lambda}[\kappa\theta_1-\theta_2,\theta]\right)\right]\Phi
\nn
+\sum_{\alpha=1}^{\infty}\frac{(2\chi)^{a}}{a!}\prwc\left(D_2([\theta_1,\theta]^a)\right)(2i\theta-(2-N)\mathcal{E}).\eea
We use (\ref{ref4}) and 
\bea \fl \prwc\left(\frac{-2\kappa\lambda}{1-\lambda}[\kappa\theta_1-\theta_2,\theta]\right)&=(fD_1+gD_2)\left(\frac{-2\kappa\lambda}{1-\lambda}[\kappa\theta_1-\theta_2,\theta]\right)\nn
& +\frac{-2\kappa\lambda}{1-\lambda}\left(f_1[\kappa\theta_1,\theta]+g_2[-\theta_2,\theta]\right)\nn
\label{ref6}&= \frac{2\kappa\lambda(f_1-g_2)}{1-\lambda}[\theta_2,\theta]\mbox{ modulo } \kappa\theta_1-\theta_2=0
\eea
to obtain a final expression for  (\ref{ref7})
\be  \prwc(D_2\Phi-u^2\Phi)= \frac{2\kappa\lambda(f_1-g_2)}{1-\lambda}[\theta_2,\theta]\Phi\ee
or, using the LSP (\ref{x}), 
 \be \prwc(D_2\Phi-u^2\Phi)= -\kappa\lambda(f_1-g_2)D_2\Phi\ee
and so, $\prwc(D_2\Phi-u^2\Phi)=0$ if and only if $f_1=g_2.$
Thus, we have shown that (\ref{ref14}) holds. Note that this is exactly the requirement that 
\be \prwc(\kappa \theta_1-\theta_2)=0 \mbox{ whenever } (\kappa \theta_1-\theta_2)=0.\ee
Hence, if $\theta$ is a traveling wave solution of the E-L equations (\ref{z}) and $\vec{w}_C$ is given by (\ref{vecw}) then  
$pr\vec{w}_C$ will be a symmetry of the LSP (\ref{x}) if and only if it is a symmetry of the traveling wave requirements, i.e. a symmetry of 
$(\kappa \theta_1-\theta_2)=0.$ Therefore, by Proposition \ref{propwlsp}, statement (\ref{1twlsp}) is equivalent to statement (\ref{2twlsp}) which is equivalent to statement (\ref{3twlsp}). \qed

In particular, what  Proposition \ref{proptwlsp} shows is that for a general conformal symmetry of the E-L equations (\ref{z}), the immersion function $\mathcal{F}$ given by (\ref{fptvlsp}) does not have tangent vectors given by (\ref{dftvlsp}) and so is generally not a Fokas-Gel'fand immersion function generated by conformal symmetries of the  $\mathbb{C}P^{N-1}$ sigma model defined on Minkowski space.  On the other hand, if  $f=ax^1+b$, $g=ax^2+c$ then the surface given in Proposition \ref{proptwlsp}  is
\be \fl \qquad  F=\Phi^\dagger(fu^1+gu^2)\Phi=-2\Phi^\dagger\left(\frac{ax^1+b}{1+\lambda}[\theta_1,\theta]+\frac{ax^2+c}{1-\lambda}[\theta_2,\theta]\right)\Phi \ee
which differs from the surface 
\bea \fl \mathcal{F}=\Phi^\dagger \prwc \Phi\nn
 \fl\qquad=\Phi^\dagger \left(-2(ax^1+b)-2\kappa (ax^2+c)+2a(\frac{\lambda x^1}{1+\lambda}-\frac{\kappa \lambda x^2}{1-\lambda}\right)[\theta_1,\theta]\Phi\nn
\fl\qquad=\Phi^\dagger\left(\frac{-2(ax^1+b)}{1+\lambda}[\theta_1,\theta]-\frac{2(ax^2+c)}{1-\lambda}[\theta_2,\theta]\right)\Phi-(\frac{2b\lambda}{1+\lambda}+\frac{2c\kappa\lambda}{1-\lambda})\Phi^\dagger[\theta_1,\theta]\Phi \nonumber
\eea
by a constant matrix 
\[ F-\mathcal{F}=(\frac{2b\lambda}{1+\lambda}+\frac{2c\kappa\lambda}{1-\lambda})\Phi^\dagger[\theta_1,\theta]\Phi.\]
In particular they are both Fokas-Gel'fand immersion functions generated by conformal symmetries of the $\mathbb{C}P^{N-1}$ sigma model defined on Minkowski space.

Thus, we have shown by counter-example that for an arbitrary generalized symmetry of the E-L equations (\ref{z}), the surface defined by (\ref{fptvlsp}) does not have  tangent vectors given by (\ref{dftvlsp}) and so is not a Fokas-Gel'fand immersion function. Equivalently the infinitesimal deformation (\ref{inftvlsp}) is not in general a symmetry of the E-L equations (\ref{z}) together with its  LSP (\ref{x}). In the next section, we will show that in the case of finite action solutions of the $\mathbb{C}P^{N-1}$ sigma model defined on Euclidean space these results do hold for arbitrary conformal transformations. 

\subsection{\textbf{$\CP$ sigma model defined on Euclidean space}} 
In this section, we consider the  $\mathbb{C}P^{N-1}$ sigma model defined on two-dimension Euclidean and, in particular, finite action solutions defined on the extended complex plane for which a complete set of solutions can be given via raising and lowering operators \cite{ BorchersGarber,DinZak1980Gen, GoldGrund2010,  Sasaki1983}. Furthermore, for this class of solutions, the wave functions $\Phi$ can be given explicitly in terms of the set of orthogonal projectors \cite{GoldGrund2010, GoldGrund2009}. Here, we will prove that a conformal symmetry of the E-L equations (\ref{EL}) is also a symmetry of the LSP (\ref{cplsp}) and so the surface given in  Proposition \ref{propconf} coincides with the surface given in Proposition \ref{propwlsp}. 

Recall \cite{GoldGrund2010, GrundPost2010}, that if $P$ is any finite action solution of the $\mathbb{C}P^{N-1}$ sigma model defined on the extended complex Euclidean plane there exists a holomorphic projector $P_0$ such that $P=\Pi_+^kP_0$ where the creation and annihilations operators are defined respectively as 
\be \label{ann} \Pi_+P=\frac{D_1PPD_2P}{tr(D_1P P D_2P)}, \qquad \Pi_-P=\frac{D_2PPD_1P}{tr(D_2P P D_1P)}.\ee
Furthermore,  the set of rank-one projectors 
\be \Lambda=\label{span}\lbrace P_0=\Pi_-^kP, \Pi_-^{k-1}P, ..., \Pi_-P, P, \Pi_+P,..., \Pi_+^{N-k-1}P\rbrace\ee
are mutually orthogonal and their images span $\mathbb{C}^N$ and $\Pi_\pm$ are contracting operators, i.e. there is some $N$ so that $\Pi_{\pm}^NP=0$ \cite{GoldGrund2010, GrundPost2010}. Finally, as was shown in \cite{ GoldGrund2010, Zakbook}, the solutions for the wave function $\Phi$ of the LSP (\ref{cplsp}) associated with the E-L equations (\ref{EL}) are given for $P=\Pi_+^kP_0$ as 
\be\label{phik} \Phi=\left(\mathbb{I}+\frac{4\lambda}{(1-\lambda)^2}\sum_{j=1}^{k}\Pi_-^jP-\frac{2}{1+\lambda}P\right).\ee  
We can simplify this notation by recalling that $\Pi_\pm$ are contracting operators and in particular $\Pi_-^{k+1}P=0$ \cite{GrundPost2010} and so it is possible to drop the dependence of $\Phi$ on $k$ and say that if $P$ is any rank-one Hermitian projector which is a finite action solution of the  $\mathbb{C}P^{N-1}$ sigma model defined on the extended complex Euclidean plane, $\Phi$ given by 
\be \label{phi} \Phi=\left(\mathbb{I}+\frac{4\lambda}{(1-\lambda)^2}\sum_{j=1}^{\infty}\Pi_-^jP-\frac{2}{1+\lambda}P\right).\ee  
is a solution of the LSP (\ref{cplsp}). Note that $\Phi$ depends on $P$ and its derivatives, i.e. $\Phi(P, D_JP)$, so it is consistent to apply the prolongation of a generalized vector field in evolutionary form on the wave function. 

Again, these results can be rewritten in terms of the matrix function  $\theta\in su(N)$ with $P=\mathcal{E}-i\theta,$ where $\mathcal{E}=\mathbb{I}/N.$ In this notation, 
\bea \Pi_+P&=\frac{D_1PPD_2P}{tr(D_1P P D_2P)}=\frac{\theta_1(\mathcal{E}-i\theta)\theta_2}{tr(\theta_1(\mathcal{E}-i\theta)\theta_2)},\\
 \Pi_-P&=\frac{D_2PPD_2P}{tr(D_2P P D_2P)}=\frac{\theta_2(\mathcal{E}-i\theta)\theta_1}{tr(\theta_2(\mathcal{E}-i\theta)\theta_1)}.
\eea
 Denoting $\Pi_\pm(P)=\Pi_\pm(\mathcal{E}-i\theta),$  the wave function $\Phi$ (\ref{phi}) becomes
\be \label{phit}  \Phi=\mathbb{I}+\frac{4\lambda}{(1-\lambda)^2}\sum_{j=1}^{\infty}\Pi_-^j\left(\mathcal{E}-i\theta\right)-\frac{2}{1+\lambda}\left(\mathcal{E}-i\theta\right).\ee

We claim that a conformal transformation on $\theta$ induces a conformal transformation on each of the orthogonal projectors $\Pi_{\pm}^k\left(\mathcal{E}-i\theta\right)$ and consequently on the wave function $\Phi.$
\begin{proposition}\label{propconp} 
 Given a generalized vector field $\vec{w}_C$ associated with a conformal symmetry transformation.
 The action of $\prwc$ on $\Pi_\pm^k\left(\mathcal{E}-i\theta\right)$ is 
\be  \prwc\left(\Pi_\pm^k(\mathcal{E}-i\theta\right)=fD_1\Pi_\pm^k\left(\mathcal{E}-i\theta\right)+gD_2\Pi_\pm^k\left(\mathcal{E}-i\theta\right).\ee
That is, the infinitesimal deformation 
\[\theta'=\theta+\epsilon(f\theta_1+g\theta_2)\] induces the infinitesimal deformation 
\bea  \left(\Pi_\pm^k\left(\mathcal{E}-i\theta\right)\right)'&=&\Pi_\pm^k\left(\mathcal{E}-i\theta\right)+\epsilon\left(fD_1 \Pi_\pm^k\left(\mathcal{E}-i\theta\right)+gD_2\Pi_\pm^k\left(\mathcal{E}-i\theta\right)\right)\nonumber.\eea

\end{proposition}
\textbf{Proof:} We prove this by induction for the $\Pi_-$ case. That is, for $\vec{w}_C$ given by (\ref{vecw}),  we shall prove that for all $k=0,1,2\ldots,$ 
\be \label{indg}  \prwc\left(\Pi_-^k(\mathcal{E}-i\theta)\right)=fD_1\Pi_-^k\left(\mathcal{E}-i\theta\right)+gD_2\Pi_-^k\left(\mathcal{E}-i\theta\right)\ee
and the proof for  $\Pi_+$ will follow by symmetry in the variables $x^1$ and $x^2.$ 
When $k=0,$ we can directly observe that 
\[\prwc\left(\mathcal{E}-i\theta\right)=-i(f\theta_1+g\theta_2)=fD_1(\mathcal{E}-i\theta)+gD_2(\mathcal{E}-i\theta)\]
so the identity (\ref{indg}) holds for $k=0.$
For the induction step, we assume that (\ref{indg}) holds for  $k$ and show that it holds for $k+1.$ 
Let us define new independent variables $\phi=\phi^je_j\in su(N)$ 
\be\label{defphi} \phi\equiv i\left(\mathcal{E}-\Pi_-^k\left(\mathcal{E}-i\theta\right)\right),\qquad \Pi_-^k\left(\mathcal{E}-i\theta\right)=\mathcal{E}-i\phi \ee
and generalized vector field $\vec{z}_C$ as
\be \vec{z}_C=\left(f D_1\phi_1^j+gD_2\phi^j\right)\frac{\partial}{\partial \phi^j}\ee
with  characteristic vector $\Q _j=f D_1\phi^j+gD_\phi^j$.
The dependent variables $\theta^j(x^1,x^2)$ and $\phi^j(x^1,x^2)$ are related via (\ref{defphi}) and its inverse
\be \theta=-i(\mathcal{E}-P)=-i(\mathcal{E}-\Pi_+^k(\Pi_-^k P))=-i(\mathcal{E}-\Pi_+^k(\mathcal{E}-i\phi)).\ee

 As before with matrix function $\theta,$ the action of the raising and lowering operators action on $\mathcal{E}-i\phi$ is given by
\be \Pi_-(\mathcal{E}-i\phi)=\frac{D_2\phi (\mathcal{E}-i\phi)D_1\phi}{tr(D_2\phi(\mathcal{E}-i\phi) D_1\phi)}.\ee
Then the induction assumption (\ref{indg}) becomes 
\bea\label{indkp} \prwc \left(\mathcal{E}-i\phi\right)&=&fD_1\left(\mathcal{E}-i\phi\right)+gD_2\left(\mathcal{E}-i\phi\right)\nonumber\\
&=&pr\vec{z}_C\left(\mathcal{E}-i\phi\right).\eea 
Using this form of the induction assumption (\ref{indkp}), we compute
\bea\label{ref9}\fl \prwc\left(\Pi_-^{k+1}\left(\mathcal{E}-i\theta\right)\right)=\prwc\left(\Pi_-\left(\mathcal{E}-i\phi\right)\right)\\
 \fl \quad=\frac{\prwc (D_2\phi (\mathcal{E}-i\phi)D_1\phi)}{tr(D_2\phi(\mathcal{E}-i\phi) D_1\phi)}-\frac{D_2\phi (\mathcal{E}-i\phi)D_1\phi}{tr(D_2\phi(\mathcal{E}-i\phi) D_1\phi)^2}tr(\prwc (D_2\phi (\mathcal{E}-i\phi)D_1\phi)\nonumber.
\eea
The fact the the prolongation of a generalized vector field commutes with the total derivatives gives
\bea \fl  \prwc ( D_2\phi (\mathcal{E}-i\phi)D_1\phi)=\\
 \fl \qquad D_2(\prwc\phi) (\mathcal{E}-i\phi)D_1\phi+ D_2\phi \prwc(\mathcal{E}-i\phi)D_1\phi+ D_2\phi (\mathcal{E}-i\phi)D_1(\prwc\phi)\nonumber.\eea
which can be written as, 
\bea \fl \prwc ( D_2\phi (\mathcal{E}-i\phi)D_1\phi)=\\
 \fl \qquad D_2(pr\vec{z}_C\phi) (\mathcal{E}-i\phi)D_1\phi+ D_2\phi pr\vec{z}_C(\mathcal{E}-i\phi)D_1\phi+ D_2\phi (\mathcal{E}-i\phi)D_1(pr\vec{z}_C \phi)\nonumber\\
\fl \qquad =pr\vec{z}_C( D_2\phi (\mathcal{E}-i\phi) D_1\phi),
\eea
where the induction assumption written in terms of $\phi$ (\ref{indkp}) was used to switch  $\vec{w}_C$ with $\vec{z}_C.$
Thus, (\ref{ref9}) can be rewritten as 
\be \prwc(\Pi_-^{k+1}(\mathcal{E}+i\theta))=pr\vec{z}_C\Pi_-(\mathcal{E}+i\phi)\ee
and it remains to show that 
\be pr\vec{z}_C\Pi_-(\mathcal{E}+i\phi)=fD_1(\Pi_-(\mathcal{E}+i\phi)+gD_2(\Pi_-(\mathcal{E}+i\phi))\ee
in order to prove (\ref{indg}) for $k+1.$ The prolongation of the vector field $\vec{z}_C$ is 
\[ pr\vec{z}_C=fD_1+gD_2+\sum_{n>0,J}(f_{n}\phi^j_{J,1}+g_{n}\phi^j_{J,2})\frac{\partial}{\partial \phi^j_{J,n}}\]
and so 
\be\fl \label{ref10} pr\vec{z}_C \Pi_-(\mathcal{E}+i\phi)=\left(fD_1+gD_2+f_1\phi^j_{1}\frac{\partial}{\partial \phi^j_{1}}+g_2\phi^j_{2}\frac{\partial}{\partial \phi^j_{2}}\right)\Pi_-(\mathcal{E}+i\phi).\ee
However, the last two terms vanish due to the form of the lowering operator, $\Pi_-,$ i.e. 
\bea \fl \phi^j_{1}\frac{\partial}{\partial \phi^j_{1}}\Pi_-(\mathcal{E}+i\phi)&=&\phi^j_{1}\frac{\partial}{\partial \phi^j_{1}}\frac{(\phi^l_1e_l)(\mathcal{E}-i\phi^me_m)(\phi^n_2e_n)}{tr\left((\phi^l_1e_l)(\mathcal{E}-i\phi^me_m)(\phi^n_2e_n)\right)}\nn
\fl  &=&\frac{(\phi^l_1e_l)(\mathcal{E}-i\phi^me_m)(\phi^n_2e_n)}{tr\left((\phi^l_1e_l)(\mathcal{E}-i\phi^me_m)(\phi^n_2e_n)\right)}\nn
&&-\frac{(\phi^l_1e_l)(\mathcal{E}-i\phi^me_m)(\phi^n_2e_n)}{tr\left((\phi^l_1e_l)(\mathcal{E}-i\phi^me_m)(\phi^n_2e_n)\right)^2}tr\left((\phi^l_1e_l)(\mathcal{E}-i\phi^me_m)(\phi^n_2e_n)\right)=0\nonumber.
\eea
A similar computation gives 
\be \phi^j_{2}\frac{\partial}{\partial \phi^j_{2}}\Pi_-(\mathcal{E}+i\phi)=0.\ee
Thus, (\ref{ref10}) becomes 
\be  pr\vec{z}_C \Pi_-(\mathcal{E}+i\phi)=fD_1\Pi_-(\mathcal{E}+i\phi))+gD_2\Pi_-(\mathcal{E}+i\phi)\ee
and we have proved that 
\bea \prwc(\Pi_-^{k+1}(\mathcal{E}+i\theta))&=fD_1\Pi_-(\mathcal{E}+i\phi)+gD_2\Pi_-(\mathcal{E}+i\phi)\nn
&=fD_1\Pi_-^{k+1}(\mathcal{E}+i\theta)+gD_2\Pi_-^{k+1}(\mathcal{E}+i\theta)\eea
which proves the induction. \qed
This proposition shows that  a conformal transformation in $\theta$ induces a conformal transformation on the entire spanning set of orthogonal projectors $\Lambda$. Furthermore, since $\Phi$ is linear in the basis $\Lambda$, a conformal transformation in $\theta$ also induces a conformal transformation in $\Phi$ and we have the following corollary. 
\begin{corollary}\label{coldphi}
 For any finite action solution of the $\mathbb{C}P^{N-1}$ sigma model defined on the extended complex Euclidean plane given in terms of rank-one Hermitian projector $P=(\mathcal{E}-i\theta),$ the action of a conformal transformation in $\theta$ induces a conformal transformation on the wave functions $\Phi$ which satisfy the LSP (\ref{cplsp}) associated with the E-L equations (\ref{EL}).  That is, for a generalized vector field $\vec{w}_C$ of the form (\ref{vecw}), its action on $\Phi$ is  
\be\label{prphi} \prwc \Phi =fD_1\Phi+gD_2\Phi.\ee

\end{corollary}
 \textbf{Proof:} From the previous lemma, the action of $\prwc$ on $\Pi_-^k(\mathcal{E}-i\theta)$ is given by (\ref{indg}). Furthermore, for any finite action solution of the E-L equations (\ref{EL}) the wave function $\Phi$ which solves  the LSP (\ref{cplsp}) is given by (\ref{phit}) and in particular is a linear combination of the lowering operators $\Pi_-^k(\mathcal{E}-i\theta).$ Thus, it is straightforward to see that (\ref{prphi}) holds. \qed

We can then use this corollary to show that for any finite action solution of the  $\mathbb{C}P^{N-1}$ sigma model defined on the extended complex Euclidean plane, the generalized vector field associated with a conformal transformation is a conformal symmetry of the LSP (\ref{cplsp}). 
We prove the results in the following proposition. 

\begin{proposition}\label{propeuc}
 Let $P=(\mathcal{E}-i\theta)$ be a finite action solution of the E-L equations (\ref{EL}) defined on the extended complex plane and $\Phi$ a solution of the associated LSP (\ref{cplsp}). Let $\vec{w}_C$ be a  generalized vector field  associated with a conformal symmetry transformation given as in (\ref{vecw}). Then the following statements hold:
\begin{enumerate}
\item\label{1euc} The generalized vector field $\vec{w}_C$ is a generalized symmetry of the LSP (\ref{cplsp}). That is 
\be \label{prlsp} \prwc(D_\alpha\Phi-u^\alpha\Phi)=0 \mbox{ whenever } D_\alpha\Phi-u^\alpha\Phi=0.\ee
\item \label{2euc} The $su(N)$-valued immersion function 
\be \label{feuc} F=\Phi^\dagger \prwc \Phi\ee
has tangent vectors given by 
\be \label{dfeuc}D_\alpha F=\Phi^\dagger \prwc u^\alpha \Phi, \qquad \alpha=1,2\ee
and so $F$ is a Fokas-Gel'fand immersion function generated by conformal symmetries of the  $\mathbb{C}P^{N-1}$ sigma model defined on Euclidean space. 
\item \label{3euc} The infinitesimal deformation 
\be \label{infeuc} \left( \ba{c} u^{1} \\ u^{2}\\ \Phi \ea \right) \rightarrow \left( \ba{c} u^{1} \\ u^{2}\\ \Phi \ea \right)+\epsilon \left( \ba{c} pr\vec{w}_Cu^1 \\ pr\vec{w}_Cu^2\\  pr\vec{w}_C \Phi \ea \right)\ee
gives a generalized infinitesimal symmetry of the system of the E-L equations (\ref{EL}) together with its associated  LSP (\ref{cplsp}).
\end{enumerate}
\end{proposition}
\textbf{Proof:} Proposition \ref{propwlsp} implies that statements (\ref{2euc}) and (\ref{3euc}) hold if and only if (\ref{1euc}) holds. Thus, to prove the proposition, we need only show that 
\[ \prwc(D_\alpha\Phi-u^\alpha\Phi)=0 \mbox{ whenever } D_\alpha\Phi-u^\alpha\Phi=0.\]
We use the fact that the prolongation of an evolutionary vector field commutes with the total derivatives (see Appendix) and (\ref{prphi}) from Corollary \ref{coldphi}
to compute
\bea \fl
\label{ref11} \prwc(D_\alpha\Phi-u^\alpha\Phi)&=D_\alpha(fD_1\Phi+gD_2\Phi)-\prwc(u^\alpha)\Phi-u^\alpha\prwc\Phi.
\eea
The action of $\prwc u^\alpha$ was given in (\ref{prwu1}) and (\ref{prwu2}) and we use this to show that, whenever $D_\alpha\Phi-u^\alpha\Phi=0,$ the requirement (\ref{ref11}) holds for $\alpha=1$ since
\bea \fl
\prwc(D_1\Phi-u^1\Phi)&=D_1(fu^1\Phi+gu^2\Phi)-D_1(fu^1)\Phi-gD_2(u^1)\Phi-u^1(fu^1+gu^2)\Phi\nn
&=g_2(D_1u^2-D_2u^1+[u^2,u^1])\Phi=0,\nonumber \eea
where the final equality is from E-L equations (\ref{c}). 
Similarly, for $\alpha=2,$ the requirement (\ref{ref11}) holds since 
\[\prwc(D_2\Phi-u^2\Phi)=f_1(D_2u^1-D_1u^2+[u^1,u^2])\Phi=0.\]
and so we have shown that $\vec{w}_\Q $ is a generalized symmetry of the LSP (\ref{cplsp}) and  therefore statement (\ref{1euc}) is true and we have proved the proposition. \qed 

To conclude, in this section we have shown that, for $\mathbb{C}P^{N-1}$ sigma models, a generalized symmetry of the E-L equations (\ref{EL}) is not always a generalized symmetry of the associated LSP (\ref{cplsp}) and so the infinitesimal deformation given by (\ref{infwt})  is not always a symmetry of the the  LSP (\ref{cplsp}). Similarly, for an arbitrary conformal symmetry of the E-L equations (\ref{EL}),  the $su(N)$-valued immersion function given by 
\be F=\Phi^{-1}\left(fu^1+gu^2\right)\Phi\in su(N)\ee
is a Fokas-Gel'fand immersion function generated by the generalized vector field $\vec{w}_C$, i.e. the immersion function $F$ has tangent vectors \be \label{dff} D_\alpha=\Phi^{-1}pr\vec{w}_Cu^\alpha \Phi.\ee Conversely, the immersion function $\mathcal{F}$ defined as 
\be \mathcal{F}=\Phi^{-1}pr\vec{w}_C\Phi \in su(N)\ee 
is not generally a Fokas-Gel'fand immersion for traveling wave solutions of the $\mathbb{C}P^{N-1}$ sigma model defined on Minkowski space but is such an immersion for finite action solutions of the $\mathbb{C}P^{N-1}$ sigma model defined on the extended complex Euclidean plane.

\section{Conclusion}

In this paper, we perform a group theoretical analysis of the Fokas-Gel'fand formula for the immersion of 2D surfaces in Lie algebras using the formalism of generalized vector fields and their prolongation structure. We provide the necessary and sufficient conditions for the existence of such surfaces in terms of the invariance criterion for generalized symmetries. The most important advantage of the presented approach is that it provides an efficient tool for the systematic construction and investigation of 2D surfaces immersed in a multi-dimensional space proceeding directly from the integrable model under consideration. We have also used this approach to consider the expression given in \cite{FGFL} for the explicit integration of the $\g$-valued immersion function, up to an appropriate constant of integration, and have derived the necessary and sufficient conditions for this reformulated expression to hold. 

These theoretical results are then illustrated by making use of  the Fokas-Gel'fand formula to construct surfaces immersed in $su(N)$ associated with the $\mathbb{C}P^{N-1}$ sigma model defined on two-dimensional Minkowski or Euclidean space. In particular, we show that the sufficient conditions for the integration of the Fokas-Gel'fand immersion function are not satisfied for arbitrary conformal symmetries of traveling wave solutions of the $\mathbb{C}P^{N-1}$ sigma model defined on Minkowski space but are identically satisfied  for arbitrary conformal symmetries of finite action solutions of the  $\mathbb{C}P^{N-1}$ sigma model defined on Euclidean space. 

This research could be expanded in several directions to further our understanding of the $\g$-valued immersion function and the geometric characteristics of the induced 2D surfaces. One possible direction of investigation is to systematically analyze the surfaces obtained from  the Fokas-Gel'fand immersion formula as applied to  the $\mathbb{C}P^{N-1}$ sigma model and to compare these surfaces with those obtained through other methods \cite{GrundSnobl2006, GSZ2005,  GrunYurd2009}. In particular, it was shown in \cite{GoldGrund2009} that, for the $\mathbb{C}P^{N-1}$ sigma model defined on Euclidean space, the Sym-Tafel formula is equivalent to the generalized Weierstrass   formula for immersion  and creation and annihilation operators were provided for a sequence of surfaces associated with the set of orthogonal projectors spanning $\mathbb{C}^N.$ It is then natural to ask how the Fokas-Gel'fand immersion, as an extension of the Sym-Tafel formula, fits into this framework and if analogous recurrence operators can be constructed. Similarly, as in \cite{GrundPost2010}, many geometric characteristics of the surfaces given by the generalized Weierstrass formula for immersion (including fundamental forms of surfaces as well as the relations between them as expressed in the Gauss-Weingarten and Gauss-Codazzi-Ricci equations, Gaussian curvature, mean curvature vector, Willmore functional and Euler-Poincar\'e characters) were expressed in terms of physical quantities of the $\mathbb{C}P^{N-1}$ sigma model (including the Euler-Lagrange equations, Lagrangian density, action and topological charge). It would be interesting to understand how the deformation of the Sym-Tafel formula, defined using the Fokas-Gel'fand formula, affects these geometric characteristics and how they are related to deformations of the physical model under consideration. 

Finally, these surfaces and their deformations have many interesting potential applications. Here we name just a few of them. In physics, potential applications include quantum field theory and string theory (both bosonic \cite{Polchbook} and superstring \cite{BGPR1994}), statistical physics \cite{McCoyWubook}, gauge field theory \cite{Amitbook} phase transitions (e.g growth of crystals, dynamics of vortex sheets, surface waves etc \cite{DGZbook,NPWbook}) and fluid dynamics (e.g. the motion of boundaries between regions of different densities and viscosities \cite{CJZbook,Saframbook}). In biochemistry and biology, surfaces and their deformations have played a fundamental role in the study of many nonlinear phenomena including the study of biological membranes and vesicles, for example long protein molecules \cite{Davbook,Ou-YLuiXie1999} and the Canham-Helfrich membrane model \cite{Landolffi2003}).  In mathematics, the construction of surfaces associated with integrable models can be applied to the study of isomonodromy deformations of surfaces and Painlev\'e type equations \cite{BobEit2000}. These applications and  further theoretical issues will be explored in our future works. 

\ack{}This work was supported by a research grant from NSERC of Canada. S. Post acknowledges a postdoctoral fellowship awarded by the Laboratory of Mathematical Physics of the Centre de Recherches Math\'ematiques, Universit\'e de Montr\'eal. 

\section*{Appendix}
\appendix
\setcounter{section}{1}
Since it is used repeatedly throughout the proofs in this paper, we include the proof of Lemma 5.12 from p. 300 of the book by P J Olver \cite{Olver}.
 We check that the prolongation of a general evolutionary vector field $\vec{w}_\Q $  commutes with the total derivative $D_\alpha.$
The form of $ D_\alpha( pr\vec{w}_\Q )$ and $ pr\vec{w}_\Q (D_\alpha)$ can be obtained by direct computations,  
\bea \label{a} \fl D_\alpha( pr\vec{w}_\Q )&=&D_{J',\alpha}\Q _n\frac{\partial}{\partial \theta^n _{J'}}+D_{J'}\Q _n\frac{\partial}{\partial x^i \partial \theta^n_{J'}}+D_{J'}\Q _n\theta^n_{J,\alpha}\frac{\partial}{\partial \theta^n_J\partial \theta^n_{J'}}\\
\fl pr\vec{w}_\Q (D_\alpha)&=&D_{J'}\Q _n\frac{\partial}{\partial \theta^n_{J'}}\left(\frac{\partial}{\partial x^\alpha}+\theta^n_{J,i}\frac{\partial}{\partial \theta^n_J}\right)\nonumber\\
\fl \label{b}  &=&D_{J,\alpha}\Q _n\frac{\partial}{\partial \theta^n _{J}}+D_{J'}\Q _n\frac{\partial}{\partial x^i \partial \theta^n_{J'}}+D_{J'}\Q _n\theta^n_{J,\alpha}\frac{\partial}{\partial \theta^n_J\partial \theta^n_{J'}}.\eea
 If we compare the two equations above (\ref{a}) and (\ref{b}), we observe that they coincide so long as 
 \[D_{J,\alpha}\Q _n\frac{\partial}{\partial \theta^n _{J}}=D_{J',\alpha}\Q _n\frac{\partial}{\partial \theta^n _{J'}}.\]
 But, since  we are using the convention that repeated indices are summed over, these are equal and so we have proved 
\be\label{A1}[D_\alpha, pr\vec{w}_\Q ]=0.\ee

\end{document}